\documentclass{article}%
\usepackage{amssymb}
\usepackage{graphicx}%
\usepackage{amsmath}%
\usepackage{amsfonts}

\begin{document}

\begin{center}
$\bigskip$

${\LARGE RELATIVE}$ ${\LARGE PHASE}$ ${\LARGE STATES}$ ${\LARGE IN}$
${\LARGE QUANTUM}{\small -}{\LARGE ATOM}$ ${\LARGE OPTICS\bigskip}$

B J Dalton${\LARGE \bigskip}$

\textit{ARC Centre for Quantum-Atom Optics}

and

\textit{Centre for Atom Optics and Ultrafast Spectroscopy, Swinburne
University of Technology, }

\textit{Melbourne, Victoria 3122, Australia\bigskip}

\bigskip
\end{center}

Email: bdalton@swin.edu.au

\begin{center}
\pagebreak
\end{center}

\subsection{Abstract}

Relative phase is treated as a physical quantity for two mode systems in
quantum atom optics, adapting the Pegg-Barnett treatment of quantum optical
phase to define a linear Hermitian relative phase operator via first
introducing a complete orthonormal set of relative phase eigenstates. These
states are contrasted with other so-called phase states. Other approaches to
treating phase and previous attempts to find a Hermitian phase operator are
discussed. The relative phase eigenstate has maximal two mode entanglement, it
is a fragmented state with its Bloch vector lying inside the Bloch sphere and
is highly spin squeezed. The relative phase states are applied to describing
interferometry experiments with Bose-Einstein condensates (BEC), particularly
in the context of a proposed Heisenberg limited interferometry experiment. For
a relative phase eigenstate the fractional fluctuation in one spin operator
component perpendicular to the Bloch vector is essentially only of order
$1/N$, so if such a highly spin squeezed state could be prepared it may be
useful for Heisenberg limited interferometry. An approach for preparing a BEC
in a state close to a relative phase state is suggested, based on
adiabatically changing parameters in the Josephson Hamiltonian starting from a
suitable energy eigenstate in the Rabi regime. \bigskip

\begin{center}
\bigskip

\pagebreak
\end{center}

\section{Introduction}

Studies of phase dependent phenomena in Bose-Einstein condensates (BEC) are
hindered because phase has at least three different meanings. A similar
situation applies in quantum optics \cite{Barnett93a}. In a first approach,
phase is regarded as a \textit{physical property} of the system
\cite{Barnett89a} and is represented via a linear Hermitian operator that
applies for all states and for which phase is the (real) eigenvalue. In a
second treatment, the state is represented in a \textit{phase space
}\cite{Schleich92a}, \cite{Smithey93a} by a quasi-distribution function, and a
complex phase is used to specify points in this space. In a third method, an
\textit{operational }approach \cite{Shapiro89a}, \cite{Noh91a} emphasises an
aparatus involved in measurements on the system and phase then refers to a
feature of this measurement aparatus, such as the phase of a classical
oscillator field interacting with the system. The dependence of system
behaviour on phase in these three approaches are different in general, making
comparison difficult because the meaning of "phase" is not the same. In a
simple example presented in \cite{Barnett93a}, it is shown that no quantum
state leads to a \textit{uniform} phase dependence for \textit{all} three
different meanings for phase. It is important therefore to recognise that
phase is different in the three approaches. Choosing which approach to use is
somewhat a matter of personal preference, but from a fundamental point of view
treating phase in the same way as other physical quantities would be
preferable - if possible. The operational phase approach has the disadvantage
of not linking phase to any intrinsic property of the system, and is dependent
on the choice of aparatus. This generally involves some sort of homodyne
system, but various combinations of beam splitters, phase shifters, vacuum
input ports, detectors etc can also be involved. The phase space approach
involves a complex phase which cannot be a measured value for any physical
quantity and is dependent on the choice of distribution function used to
describe the state. This could be Wigner W, Glauber-Sudarshan P, Husimi Q, .or
other distributions. In fact the complex phase is often the eigenvalue of a
non-Hermitian annihilation operator, which cannot represent a physical
quantity. In neither of these two approaches is there any unique or compelling
choice for defining phase. Thus the introduction of phase via eigenvalues of a
linear Hermitian operator associated with the quantum system is the most
objective approach \cite{Barnett93a} because it is not dependent on any
particular way of specifying the state nor on any particular measurement
system. This is not to claim that some ways of specifying the state are not
more useful than others, nor is it intended to trivialise the difficult issue
of measuring the probablity distribution for the measurable values of phase
regarded as a physical property. Nor does it prove easy to find a suitable
Hermitian operator to represent phase. However such an operator can now be
defined both for quantum optical systems and Bose-Einstein condensates
following the approach of Pegg and Barnett \cite{Barnett89a} that was
originally applied to the quantum optics case (see \cite{Pegg97a} for a recent review).

In the present paper we begin with a brief review of progress towards finding
a Hermitian phase operator, and then the Pegg-Barnett approach is adapted to
define a relative phase operator for two mode systems via first introducing a
complete orthonormal set of relative phase eigenstates. These states are
contrasted with other so-called phase states. Interesting properties of the
relative phase eigenstates are then determined, these being entanglement,
fragmentation and spin squeezing. In the final section applications of the
relative phase states in describing BEC interferometry experiments are made
based on treatment involving the Josephson Hamiltonian, and possibilities for
preparing a BEC in a relative phase state are examined. Certain technical
results needed for the main body of the paper are covered in an Appendix,
which is available as online Supplementary Data.\medskip

\section{Hermitian phase operators}

\subsection{Early attempts}

Early attempts to find a \textit{Hermitian phase operator} for each mode by
expressing the annihilation and creation operators $\widehat{a}$, $\widehat
{a}^{\dag}$ in terms of Hermitian number $\widehat{n}_{a}$ and phase operator
$\widehat{\phi}_{a}$ via $\widehat{a}=\exp(i\widehat{\phi}_{a})\,(\widehat
{n}_{a})^{1/2}$, $\widehat{a}^{\dag}=(\widehat{n}_{a})^{1/2}\,\exp
(-i\widehat{\phi}_{a})$ with $[\widehat{\phi}_{a},\widehat{n}_{a}]=-i$
\cite{Dirac27a}, or by introducing two \textit{exponential operators}
\cite{Susskind64a}, \cite{Nieto93a} via the annihilation and creation
operators $\widehat{E^{+}}=(\widehat{n}_{a}+1)^{-1/2}\,\widehat{a}$,
$\widehat{E^{-}}=\widehat{a}^{\dag}\,(\widehat{n}_{a}+1)^{-1/2}=(\widehat
{E^{+}})^{\dag}$ were unsuccessful. For the first, the commutation rule leads
to a contradiction when matrix elements between number states are evaluated -
$(n-m)\left\langle n|\widehat{\phi}_{a}|m\right\rangle =-i\delta_{nm}$. For
the second, the introduction of a Hermitian phase operator required being able
to write $\widehat{E^{+}}=\exp(i\widehat{\phi}_{a})$ and $\widehat{E^{-}}%
=\exp(-i\widehat{\phi}_{a})$ in the form of unitary operators also failed, for
although $\widehat{E^{+}}\widehat{E^{-}}=\widehat{1}$ we have $\widehat{E^{-}%
}\widehat{E^{+}}=\widehat{1}-\left\vert 0\right\rangle \left\langle
0\right\vert $ - rather than $\widehat{1}$ as is required. Note that Hermitian
\textit{cosine} and \textit{sine operators} $\widehat{C}$, $\widehat{S}$ can
be introduced via $\widehat{E^{+}}=\widehat{C}+i\widehat{S}$, but again these
are not trigonometric functions of a Hermitian phase operator. However, the
approach of Pegg and Barnett \cite{Barnett89a} via the introduction of phase
eigenstates and then the Hermitian phase operator was successful. This
approach does require introducing a cut-off on boson numbers for each mode,
but this can be justified mathematically in terms of a sequence of Hilbert
spaces \cite{Vaccaro95a} and shown not to affect physical predictions for
finite energy fields.\smallskip

\subsection{Relative phase operator and eigenstates}

In this section we introduce \textit{relative phase} \textit{eigenstates} and
a Hermitian \textit{relative phase operator} for the case of a two mode
\textit{single component} BEC with mode annihilation operators $\widehat{a}$,
$\widehat{b}$ and spatial mode functions $\phi_{a}(\mathbf{r})$, $\phi
_{b}(\mathbf{r})$ using a modification of the Pegg-Barnett approach for single
modes. From the Fock state orthonormal basis states $\left\vert n_{a}%
\right\rangle $, $\left\vert n_{b}\right\rangle $ involving $n_{a}$, $n_{b}$
bosons in the modes a complete orthonormal set of relative phase eigenstates
$\left\vert \theta_{p}\right\rangle $ for the $N=n_{a}+n_{b}$ boson system are
then defined via%
\begin{equation}
\left\vert \theta_{p}\right\rangle =\frac{1}{\sqrt{N+1}}\sum_{k=-N/2}%
^{N/2}\exp(ik\theta_{p})\left\vert N/2-k\right\rangle _{a}\left\vert
N/2+k\right\rangle _{b}\label{Eq.RelPhaseEigenstate}%
\end{equation}
where $\theta_{p}=p(2\pi/(N+1))$, $p=-N/2,-N/2+1,..,+N/2$ is a quasi-continuum
of $N+1$ equispaced phase eigenvalues. The Hermitian relative phase operator
is then defined as
\begin{equation}
\widehat{\Theta}=\sum_{p}\theta_{p}\left\vert \theta_{p}\right\rangle
\left\langle \theta_{p}\right\vert \label{Eq.RelPhaseOpr}%
\end{equation}
This approach has also been applied previously in \cite{Luis93a},
\cite{Luis96a}. Also an un-normalised version of
Eq.(\ref{Eq.RelPhaseEigenstate}) with phase angle $-\theta_{p}$ is introduced
in \cite{Menotti01a} (see Eqs. A3 and A4). The relative phase operator defined
in (\ref{Eq.RelPhaseOpr}) depends on the choice of modes and the total boson
number $N$, and its commutation law with the relative number operator
$\widehat{\delta n}=\frac{1}{2}(\widehat{b}^{\dag}\widehat{b}-\widehat
{a}^{\dag}\widehat{a})$ is $[\widehat{\Theta},\widehat{\delta n}]=i%
{\textstyle\sum\limits_{p\neq q}}
\left\vert \theta_{p}\right\rangle \left\langle \theta_{q}\right\vert
(-1)^{p-q}\frac{{\LARGE (\theta}_{p}{\LARGE -\theta}_{q}{\LARGE )/2}}%
{\sin{\LARGE (\theta}_{p}{\LARGE -\theta}_{q}{\LARGE )/2}}$ rather than just
$i$. The Fock state $\left\vert N/2-k\right\rangle _{a}\left\vert
N/2+k\right\rangle _{b}$ is an eigenstate of the relative number operator with
eigenvalue $k$. Note that the present approach for a two mode system defines a
\textit{relative} phase eigenstate and relative phase operator, rather than
phase eigenstates and operator for each mode. However, the relative phase
operator can be defined without requiring a cut-off on boson numbers since
there is an automatic restriction for $k$ to lie between $-N/2$ and $+N/2$.
The definition can be extended to apply to mixed state boson systems with a
range of $N$ via $\widehat{\Theta}_{T}=%
{\textstyle\sum\limits_{N}}
\widehat{\Pi}_{N}\widehat{\Theta}(N)\widehat{\Pi}_{N}$ using projectors
$\widehat{\Pi}_{N}$ onto $N$ boson states. Note that essentially the same
states can also be defined for quantum optical systems, there the bosons are
massless photons.

An approach closer to the original Pegg and Barnett method would be to define
phase operators for each mode, and then the relative phase operator would be
the difference between the separate phase operators for the two modes, and
this method is used in Ref.\cite{Barnett90a}. This approach requires
introducing a cut-off on boson numbers for each mode and special techniques
are needed to restrict the phase difference to a $2\pi$ rather than $4\pi$
interval. There are differences between this approach and that adopted here
and in \cite{Luis93a}, \cite{Luis96a}, which are discussed in \cite{Pegg95a},
\cite{Luis95a}. The approach presented here provides a more direct focus on
relative phase as a basic physical property and enables the relative phase to
automatically lie in a $2\pi$ interval.

A similar approach to that here can also be used to define relative phase
eigenstates and a Hermitian relative phase operator for a \textit{two
component} BEC where each (hyperfine) component is associated with a single
spatial mode function. This situation is again a two mode system and similar
Fock states to $\left\vert N/2-k\right\rangle _{a}\left\vert
N/2+k\right\rangle _{b}$ act as an othonormal basis, though now $\left\vert
n_{a}\right\rangle $ has $n_{a}$ bosons in a spatial mode $\phi_{a}%
(\mathbf{r})$ associated with internal (hyperfine) state $a$.\smallskip

\subsection{Pure states and quantum superpositions}

Any pure quantum state $\left\vert \Phi\right\rangle $ for the $N$ boson
system can be expanded in terms of the relative phase states as
\begin{equation}
\left\vert \Phi\right\rangle =%
{\textstyle\sum\limits_{p=-N/2}^{N/2}}
A(\theta_{p})\left\vert \theta_{p}\right\rangle
\label{Eq.PureStatePhaseStatesExpn}%
\end{equation}
and the amplitudes $A(\theta_{p})$ determine the probability $P(\theta_{p}) $
for measuring the relative phase $\theta_{p}$ via the standard expression%
\begin{equation}
P(\theta_{p})=|A(\theta_{p})|^{2}\label{Eq.ProbRelPhase}%
\end{equation}
The same state can also be expanded in terms of the relative number states as%
\begin{equation}
\left\vert \Phi\right\rangle =%
{\textstyle\sum\limits_{k=-N/2}^{N/2}}
b_{k}\left\vert N/2-k\right\rangle _{a}\left\vert N/2+k\right\rangle
_{b}\label{Eq.PureStateNumberStatesExpn}%
\end{equation}
with expansion coefficients $b_{k}$. It is then easy to see that the expansion
coefficients in terms of relative phase states and the Fock states are related
via a \textit{Fourier transform}.
\begin{equation}
A(\theta_{p})=\frac{1}{\sqrt{N+1}}%
{\textstyle\sum\limits_{k}}
\exp(-ik\theta_{p})b_{k}\qquad b_{k}=\frac{1}{\sqrt{N+1}}%
{\textstyle\sum\limits_{p}}
\exp(+ik\theta_{p})A(\theta_{p})\label{Eq.RelnCoefts}%
\end{equation}
For the relative phase state itself the expansion coefficients are $b_{k}%
=\exp(+ik\theta_{p})/\sqrt{N+1}$. The generalisation for mixed states is straightforward.

As an example of a quantum superposition we consider the state%
\begin{equation}
\left\vert \,\Phi\right\rangle =\frac{1}{\sqrt{2}}(\left\vert \,\frac
{{\small N}}{{\small 2}},-\frac{{\small N}}{{\small 2}}\right\rangle
+\left\vert \,\frac{{\small N}}{{\small 2}},+\frac{{\small N}}{{\small 2}%
}\right\rangle )\label{Eq.NOONState}%
\end{equation}
which is the so-called \textit{NOON}\ state, being a superposition of states
$\left\vert \,N,0\right\rangle $ and $\left\vert \,0,N\right\rangle $. It is
also referred to as a \textit{Schrodinger cat} state, and is an example of an
\textit{entangled} state. In the first term there are $N$ bosons in mode
$\phi_{{\small L}}$ and $0$ in mode $\phi_{{\small R}}$ and for the second the
reverse applies. For this state $b_{k}=(\delta_{k,-N/2}+\delta_{k,+N/2}%
)/\sqrt{2}$ and hence $A(\theta_{p})=\sqrt{2/(N+1)}\cos(\frac{{\small N}%
}{{\small 2}}\theta_{p})$, which gives an oscillatory probability distribution
for the relative phase with probabilities changing from $0$ to $2/(N+1)$ for
neighboring phase angles. Such oscillations would be hard to detect. On the
other hand the different NOON\ state
\begin{equation}
\left\vert \,\Phi\right\rangle =\frac{1}{\sqrt{2}}(\left\vert \,\frac
{{\small N}}{{\small 2}},-n\right\rangle +\left\vert \,\frac{{\small N}%
}{{\small 2}},+n\right\rangle )\label{Eq.NOONState2}%
\end{equation}
with $n\ll N$ and $A(\theta_{p})=\sqrt{2/(N+1)}\cos(n\theta_{p})$ would have a
central peak in the phase probability for $\theta_{p}=0$ and the first zero at
$\theta_{p}=\pm\pi/2n$, which would correspond to a relative narrow phase
probability distribution with $\Delta\theta_{p}\varpropto1/n$, if $n$ is large
enough.\smallskip

\subsection{Other phase dependent states}

Note that other authors \cite{Li09a} have defined a set of states for a two
mode BEC that depend on phase variables $\theta$, $\chi$ via the expression
\begin{equation}
\left\vert \theta,\chi\right\rangle =\frac{1}{\sqrt{N!}}(\cos\theta\cdot
\exp(-i\frac{1}{2}\chi)\widehat{a}^{\dag}+\sin\theta\cdot\exp(+i\frac{1}%
{2}\chi)\widehat{b}^{\dag})^{N}\left\vert N\right\rangle _{a}\left\vert
0\right\rangle _{b}\label{Eq.BinomialState}%
\end{equation}
which are \textit{also} referred to as \textit{phase states}. For the case
where $\theta=\pi/4$ such states have been used to define a phase $\chi$ in
BEC interferometry experiments \cite{Grond10a}, and $\chi$ is measured in
terms of the evolution time for a condensate in a double well trap when
inter-well tunneling dominates over collisional effects. In this case phase is
essentially the evolution time, which is an \textit{operational variable}
directly associated with the specific measurement process. However, states
such as $\left\vert \theta,\chi\right\rangle $ are actually \textit{binomial
states} and correspond to all bosons being in the same single particle state
$\cos\theta\cdot\exp(-i\frac{1}{2}\chi)\phi_{a}(\mathbf{r})+\sin\theta
\cdot\exp(+i\frac{1}{2}\chi)\phi_{b}(\mathbf{r})$. They are also referred to
as \textit{coherent states}. Expanding the coherent states
\begin{equation}
\left\vert \theta,\chi\right\rangle =%
{\textstyle\sum\limits_{k}}
b_{k}(\theta,\chi)\left\vert N/2-k\right\rangle _{a}\left\vert
N/2+k\right\rangle _{b}\label{Eq.BinomialStateExpn}%
\end{equation}
as a superposition of the basis states $\left\vert N/2-k\right\rangle
_{a}\left\vert N/2+k\right\rangle _{b}$, the expansion coefficients are
$b_{k}(\theta,\chi)=C_{N/2-k}^{N}\cdot(\cos\theta)^{N/2-k}\cdot(\sin
\theta)^{N/2+k}\cdot\exp(ik\chi)$ involving binomial coefficients
$C_{N/2-k}^{N}=N!/((N/2-k)!(N/2+k)!)$. The binomial states are physically
important since they describe an \textit{unfragmented} BEC \cite{Leggett01a}.
However they are not a complete orthonormal basis set for the two mode BEC.
For example there is no choice of $\theta$, $\chi$ that gives the
\textit{fragmented} state $\left\vert N/2\right\rangle _{a}\left\vert
N/2\right\rangle _{b}$ which has an occupancy for each of the two natural
orbitals (see below) of $N/2$. For the coherent state with equal probabilities
of finding a boson in each mode $\left\vert \theta=\pi/4,\chi\right\rangle $
the expansion (\ref{Eq.PureStatePhaseStatesExpn})\ in terms of relative phase
states $\left\vert \theta_{p}\right\rangle $ gives a relative phase
probability
\begin{equation}
P_{\pi/4,\chi}(\theta_{p})=\sqrt{\frac{2\pi}{N}}\sqrt{\frac{N}{N+1}}%
\exp(-\frac{N(\theta_{p}-\chi)^{2}}{2})\label{Eq.PhaseProbCoherentState}%
\end{equation}
for $N$ large. This is a Gaussian distribution centred around $\theta_{p}%
=\chi$ with a narrow width of $\Delta\theta_{p}\propto1/\sqrt{N}$. For this
coherent state the relative phase distribution corresponds to the
\textit{standard quantum limit}. Note that for large $N$ this coherent state
almost has a well-defined relative phase $\chi$, which may explain why it is
sometimes regarded as being a state with a definite relative phase. However,
they are not eigenstates of any relative phase operator.

Other authors \cite{Vogel91a} consider eigenstates $\left\vert E(\theta
)\right\rangle $ of a phase dependent \textit{quadrature operator} for each
mode $\widehat{E}(\theta)=i(\widehat{a}\exp(-i\theta)-\widehat{a}^{\dag}%
\exp(-i\theta))$ with eigenvalue $E(\theta)$, and probability distributions
given as $|\left\langle E(\theta)|\Phi\right\rangle |^{2}$ for finding the
quadrature field to have an amplitude $E(\theta)$ considered as a function of
phase variable $\theta$ determine a phase distribution for a quantum state
$\left\vert \Phi\right\rangle $ without introducing phase as an eigenvalue of
a Hermitian operator. The expansion of the quadrature eigenstates in terms of
Fock states $\left\vert n_{a}\right\rangle _{a}$ involves Hermite polynomials
and Gaussian functions of $E(\theta)$. However, the states $\left\vert
E(\theta)\right\rangle $ are non-orthogonal so the probability concept is
doubtful. This treatment of phase is really an example of the \textit{phase
space} approach.\medskip

\section{Properties of relative phase eigenstates}

The relative phase eigenstate has several interesting properties. These
include entanglement, fragmentation and spin squeezing. We deal with each in
turn.\smallskip

\subsection{Mode entanglement}

Firstly, it is a state with maximal \textit{mode entanglement} for the $a$,
$b$ sub-systems, so is of interest in quantum information \ The
\textit{entropy of entanglement} is one of the standard measures of
entanglement \cite{Amico08a} and is given by the von Neumann entropy for the
reduced density operator for either of the subsystems $a$ or $b$. Thus for the
system in pure state $\left\vert \Phi\right\rangle $ the entropy of
entanglement is
\begin{align}
S(\widehat{\rho}_{a})  & =-k_{B}Tr(\widehat{\rho}_{a}\log\widehat{\rho}%
_{a})=-k_{B}Tr(\widehat{\rho}_{b}\log\widehat{\rho}_{b})=S(\widehat{\rho}%
_{b})\nonumber\\
\widehat{\rho}_{a}  & =Tr_{b}(\left\vert \Phi\right\rangle \left\langle
\Phi\right\vert )\qquad\widehat{\rho}_{b}=Tr_{a}(\left\vert \Phi\right\rangle
\left\langle \Phi\right\vert )\label{Eq.ModeEntropy}%
\end{align}
For the relative phase eigenstate it is straightforward to show that the
entropy of entanglement is given by%
\begin{equation}
S(\widehat{\rho}_{a})=k_{B}\log(N+1)=S(\widehat{\rho}_{b}%
)\label{Eq.EntangEntropy}%
\end{equation}
which is very large. The general case of maximal mode entanglement occurs when
the amplitudes $b_{k}$ in (\ref{Eq.PureStateNumberStatesExpn}) satisfy
$|b_{k}|=1/\sqrt{N+1}$ \cite{Hines03a}, and the relative phase eigenstate is a
particular case.\smallskip

\subsection{Fragmentation}

Secondly, it is a \textit{fragmented state} \cite{Leggett01a}, since there are
two \textit{natural orbitals} with macroscopic occupancy. For large $N$ the
first order \textit{quantum correlation function} $G^{(1)}(\mathbf{r}%
,\mathbf{r}^{\prime})=\left\langle \widehat{\Psi}^{\dag}(\mathbf{r}%
)\widehat{\Psi}(\mathbf{r}^{\prime})\right\rangle $ (where $\widehat{\Psi
}^{\dag}(\mathbf{r}),\widehat{\Psi}(\mathbf{r})$ are the usual \textit{field
operators}, $\widehat{\Psi}(\mathbf{r})=\widehat{a}\phi_{a}(\mathbf{r}%
)+\widehat{b}\phi_{b}(\mathbf{r})$) is given by (see Appendix)
\begin{align}
& G^{(1)}(\mathbf{r},\mathbf{r}^{\prime})=\frac{N}{2}(\phi_{a}^{\ast
}(\mathbf{r})\phi_{a}(\mathbf{r}^{\prime})+\phi_{b}^{\ast}(\mathbf{r})\phi
_{b}(\mathbf{r}^{\prime}))\nonumber\\
& +\frac{\pi N}{8}\exp(i\theta_{p})(\phi_{a}^{\ast}(\mathbf{r})\phi
_{b}(\mathbf{r}^{\prime}))+\frac{\pi N}{8}\exp(-i\theta_{p})(\phi_{b}^{\ast
}(\mathbf{r})\phi_{a}(\mathbf{r}^{\prime}))\label{Eq.QCF}%
\end{align}
using the result that for large $N$ the sum $\sum_{k}\sqrt{(\frac{{\small N}%
}{{\small 2}}(\frac{{\small N}}{{\small 2}}+{\small 1})-k(k\pm{\small 1}%
)}/((N+1))$ is approximately $\frac{{\LARGE \pi}}{{\LARGE 8}}N$. The natural
orbitals are the eigenfunctions of the first order quantum correlation
function, and are given by $\chi_{\pm}(\mathbf{r})=(\exp(i\theta_{p}%
/2)\phi_{a}^{\ast}(\mathbf{r})\pm\exp(-i\theta_{p}/2)\phi_{b}^{\ast
}(\mathbf{r}))/\sqrt{2}$. The eigenvalues (which give the occupancies) are
$(\frac{1}{2}\pm\frac{\pi}{8})N$. Fragmented states cannot be described by a
single Gross-Pitaevskii equation, generalised mean field theories are involved
\cite{Dalton07a}, \cite{Dalton11a} involving coupled \textit{generalised
Gross-Pitaevskii equations}.\smallskip

\subsection{Spin squeezing}

Thirdly, the relative phase eigenstate is a \textit{spin squeezed state}
\cite{Kitagawa93a} in which one component of the spin angular momentum has a
Heisenberg limited fluctuation. The Schwinger spin operators are defined by
$\widehat{S}_{x}=(\widehat{b}^{\dag}\widehat{a}+\widehat{a}^{\dag}\widehat
{b})/2$, $\widehat{S}_{y}=(\widehat{b}^{\dag}\widehat{a}-\widehat{a}^{\dag
}\widehat{b})/2i$, $\widehat{S}_{z}=(\widehat{b}^{\dag}\widehat{b}-\widehat
{a}^{\dag}\widehat{a})/2$, and the \textit{Bloch vector} is defined via its
components $\left\langle \widehat{S}_{x}\right\rangle ,\left\langle
\widehat{S}_{y}\right\rangle $ and $\left\langle \widehat{S}_{z}\right\rangle
$ \cite{Jaaskelainen06a}, often in units of $N$. For large $N$ the Bloch
vector is determined to be (see Appendix)
\begin{equation}
\left\langle \widehat{S}_{x}\right\rangle \doteqdot\frac{\pi}{8}N\,\cos
\theta_{p}\qquad\qquad\qquad\left\langle \widehat{S}_{y}\right\rangle
\doteqdot-\frac{\pi}{8}N\,\sin\theta_{p}\qquad\qquad\qquad\left\langle
\widehat{S}_{z}\right\rangle =0\,\label{Eq.BlochVectorPhaseState}%
\end{equation}
using the result that for large $N$ the sum $\sum_{k}\sqrt{(\frac{{\small N}%
}{{\small 2}}(\frac{{\small N}}{{\small 2}}+{\small 1})-k(k\pm{\small 1}%
)}/((N+1))$ is approximately $\frac{{\LARGE \pi}}{{\LARGE 8}}N\doteqdot
0.3926N$. This vector is in the equatorial plane with azimuthal angle
$\phi=2\pi-\theta_{p}$, and is inside the \textit{Bloch sphere} of radius
$N/2$ - another indicator of fragmentation. For \textit{any} unfragmented
state is a coherent state $\left\vert \theta,\chi\right\rangle $ and the Bloch
vector \textit{always} lies on the Bloch sphere, the orientation being given
by polar angle $\pi-2\theta$ and azimuthal angle $2\pi-\chi$. Thus any state
for which the Bloch vector lies inside the Bloch sphere must be a fragmented
state, and the relative phase eigenstate is such a case. \smallskip

Spin operators along $(\widehat{J}_{z})$ and perpendicular $(\widehat{J}_{x}$,
$\widehat{J}_{y})$ to the Bloch vector may be defined by
\begin{equation}
\widehat{J}_{x}=\widehat{S}_{z}\qquad\widehat{J}_{y}=\widehat{S}_{x}\sin
\theta_{p}+\widehat{S}_{y}\cos\theta_{p}\qquad\widehat{J}_{z}=\widehat{S}%
_{x}\cos\theta_{p}-\widehat{S}_{y}\sin\theta_{p}\label{Eq.NewSpinOprs}%
\end{equation}
and in terms of the new spin operators
\begin{equation}
\left\langle \widehat{J}_{x}\right\rangle =0\qquad\left\langle \widehat{J}%
_{y}\right\rangle =0\qquad\left\langle \widehat{J}_{z}\right\rangle =\frac
{\pi}{8}N\approx0.392N\label{Eq.BlochVector}%
\end{equation}
\smallskip

The \textit{covariance matrix} (see Appendix) which describes the quantum
fluctuations for the spin operator components can be shown to be diagonal for
the new spin operators $\widehat{J}_{x}$, $\widehat{J}_{y}$, $\widehat{J}_{z}%
$. For large $N$ the fluctuations in the new Bloch vector components for the
relative phase eigenstate are found to be (see Appendix)
\begin{equation}
\delta\widehat{J}_{x}\approx\sqrt{1/12}N\approx0.289N\qquad\delta\widehat
{J}_{y}\approx\sqrt{\frac{1}{8}+\frac{1}{4}\ln N}.\qquad\delta\widehat{J}%
_{z}\approx\sqrt{(1/6-\pi^{2}/64)}N\approx0.112N\label{Eq.BlocVectorFlns}%
\end{equation}
where $\delta\widehat{\Omega}^{2}\equiv\left\langle (\widehat{\Omega
}-\left\langle \widehat{\Omega}\right\rangle )^{2}\right\rangle $. As
$|\left\langle \widehat{J}_{z}\right\rangle |/2\approx0.196N$ we see that for
all $N>4$ the product $\delta\widehat{J}_{x}\cdot\delta\widehat{J}_{y}%
>0.198N$, which is greater than $|\left\langle \widehat{J}_{z}\right\rangle
|/2$ consistent with the Heisenberg uncertainty principle. However, although
$\widehat{J}_{x}$ is not squeezed, the other perpendicular component
$\widehat{J}_{y}$ is highly squeezed, with a fractional fluctuation
$\delta\widehat{J}_{y}/\left\langle \widehat{J}_{z}\right\rangle $ essentially
of order $1/N$ due to the denominator $\left\langle \widehat{J}_{z}%
\right\rangle $. The numerator $\delta\widehat{J}_{y}$ is a very slowly
increasing function of $N$ - for $N$ changing from $10^{8}$ to $10^{10}$ it
only changes from $2.17$ to $2.42.$The relative phase state could be of
interest in Heisenberg limited interferometry \cite{Bouyer97a}. By contrast,
the fluctuations in the Bloch vector components for the coherent state
$\left\vert \theta,\chi\right\rangle $ are $\delta\widehat{J}_{x}\approx
\sqrt{N}$, $\delta\widehat{J}_{y}\approx\sqrt{N}$ and $\delta\widehat{J}%
_{z}\approx0$. Here the fluctuations are equal for the two components
perpendicular to the Bloch vector, so there is no squeezing. Furthermore, the
fractional fluctuation $\delta\widehat{J}_{x,y}/\left\langle \widehat{J}%
_{z}\right\rangle $ is only of order $1/\sqrt{N}$, corresponding to the
\textit{standard quantum limit} and not to the \textit{Heisenberg limit}, as
is the case for the relative phase eigenstate. \medskip

\section{Applications of relative phase eigenstates}

The relative phase eigenstate provides a useful theoretical concept for
describing interferometry experiments based on BEC. This type of application
is discussed in this Section,firstly in general terms and then for a specific
BEC interferometry proposal. However, before treating these applications the
question of whether the relative phase eigenstate can be prepared via some
sort of dynamical process will be examined.

\subsection{Creating relative phase eigenstates ?}

The energy and energy fluctuation associated with the relative phase
eigenstate are quite large. The typical two mode system such as bosons in a
double well potential is described via the Josephson Hamiltonian%
\begin{equation}
\widehat{H}=-J\widehat{S}_{x}+\delta\widehat{S}_{z}+U\widehat{S}_{z}%
^{2}\label{Eq.JosephHam}%
\end{equation}
where $J$ is the inter-well \textit{tunneling} parameter, $\delta$ describes
\textit{asymmetry} of the two wells and $U$ is the \textit{collision}
parameter. It is easy to see that the relative phase state
(\ref{Eq.RelPhaseEigenstate}) is not an energy eigenstate. The non-zero matrix
elements of the Josephson Hamiltonian between the basis states $\left\vert
N/2,k\right\rangle \equiv\left\vert N/2-k\right\rangle _{a}\left\vert
N/2+k\right\rangle _{b}$ are
\begin{align}
H_{k,k}  & =\delta k+Uk^{2}\nonumber\\
H_{k,k+1}  & =-\frac{J}{2}\sqrt{\frac{N}{2}(\frac{N}{2}+1)-(k+1)k}\nonumber\\
H_{k,k-1}  & =-\frac{J}{2}\sqrt{\frac{N}{2}(\frac{N}{2}+1)-(k-1)k}%
\label{Eq.HamMatrix}%
\end{align}
and for the relative phase state to be an energy eigenstate with energy $E$
requires
\begin{align}
E  & =\delta k+Uk^{2}-\frac{J}{2}\sqrt{\frac{N}{2}(\frac{N}{2}+1)-(k+1)k}%
\exp(+i\theta_{p})\nonumber\\
& -\frac{J}{2}\sqrt{\frac{N}{2}(\frac{N}{2}+1)-(k-1)k}\exp(-i\theta
_{p})\label{Eq.NoEnergyEigenstate}%
\end{align}
for all $k$, which is not possible. The mean energy $\left\langle \widehat
{H}\right\rangle $ is
\begin{equation}
\left\langle \widehat{H}\right\rangle \doteqdot U\frac{1}{12}N^{2}-J\cos
\theta_{p}\frac{\pi}{8}N\label{Eq.MeanEnergy}%
\end{equation}
so in the \textit{Rabi regime} \cite{Leggett01a} where $J\gg UN$ the mean
energy is approximately $-J\cos\theta_{p}\frac{{\LARGE \pi}}{{\LARGE 8}}N$,
whilst in the \textit{Fock regime} \cite{Leggett01a} where $U\gg JN$ it is
essentially $U\frac{{\LARGE 1}}{{\LARGE 12}}N^{2}$.

The \textit{variance }in the energy $\delta\widehat{H}^{2}=\left\langle
(\widehat{H}-\left\langle \widehat{H}\right\rangle )^{2}\right\rangle $ is
given by
\begin{equation}
\delta\widehat{H}^{2}\doteqdot N^{2}\times\left[
\begin{tabular}
[c]{lll}%
$UN$ & $J$ & $\delta$%
\end{tabular}
\right]  \times\left[
\begin{tabular}
[c]{lll}%
$\frac{{\Large 1}}{{\Large 180}}$ & $\frac{{\Large \pi}}{{\Large 384}}%
\cos\theta_{p}$ & $0$\\
$\frac{{\Large \pi}}{{\Large 384}}\cos\theta_{p}$ & $(\frac{{\Large 1}%
}{{\Large 6}}-\frac{{\Large \pi}^{2}}{{\Large 64}})\cos^{2}\theta_{p}$ & $0$\\
$0$ & $0$ & $\frac{{\Large 1}}{{\Large 12}}$%
\end{tabular}
\right]  \times\left[
\begin{tabular}
[c]{l}%
$UN$\\
$J$\\
$\delta$%
\end{tabular}
\right] \label{Eq.VarianceEnergy}%
\end{equation}
correct to $O(N^{2})$ (see Appendix for details). This is a \textit{quadratic
form} in the quantities $UN$, $J$ and $\delta$. That this form is positive
definite can be shown by determining the eigen values $\lambda_{1}(\theta
_{p})$, $\lambda_{2}(\theta_{p})$ and $\lambda_{3}(\theta_{p})$ of the $3x3$
matrix in Eq.(\ref{Eq.VarianceEnergy}), and explict formulae are given in the
Appendix. As expected the eigenvalues are all real and positive for all
relative phase $\theta_{p}$ (see Figure A in Appendix). It is of some interest
to consider cases where the Josephson parameters are related via
\begin{equation}
\left[
\begin{tabular}
[c]{l}%
$UN$\\
$J$\\
$\delta$%
\end{tabular}
\right]  =K\left[
\begin{tabular}
[c]{l}%
$X_{1\alpha}$\\
$X_{2\alpha}$\\
$X_{3\alpha}$%
\end{tabular}
\right] \label{Eq.SpecialJosephParam}%
\end{equation}
where $\left[
\begin{tabular}
[c]{lll}%
$X_{1\alpha}$ & $X_{2\alpha}$ & $X_{3\alpha}$%
\end{tabular}
\right]  ^{T}$ are the orthonormal column eigenvectors associated with the
eigenvalues $\lambda_{1}(\theta_{p})$, $\lambda_{2}(\theta_{p})$ and
$\lambda_{3}(\theta_{p})$ and $K$ is arbitrary. In this case an expression for
the \textit{relative energy fluctuation} can be obtained as
\begin{equation}
\frac{\sqrt{\left(  \delta\widehat{H}^{2}\right)  _{\alpha}}}{\left\vert
\left\langle \widehat{H}\right\rangle _{\alpha}\right\vert }=\frac
{\sqrt{\lambda_{\alpha}(\theta_{p})}}{\left\vert \left(  X_{1\alpha}\frac
{1}{12}-X_{2\alpha}\cos\theta_{p}\frac{\pi}{8}\right)  \right\vert
}\label{Eq.RelativeEnergyFluctnSpecialParam}%
\end{equation}
for the eigenvalues $\lambda_{1}(\theta_{p})$, $\lambda_{2}(\theta_{p})$. For
$\lambda_{3}(\theta_{p})$ we have $\left\langle \widehat{H}\right\rangle
_{3}=0$, so the relative fluctuation is undefined. In Figures 1 and 2 the
relative energy fluctuations are shown for $\lambda_{1}(\theta_{p})$,
$\lambda_{2}(\theta_{p})$ respectively.

\bigskip%
\begin{figure}
[ptb]
\begin{center}
\includegraphics[
height=3.1038in,
width=5.0548in
]%
{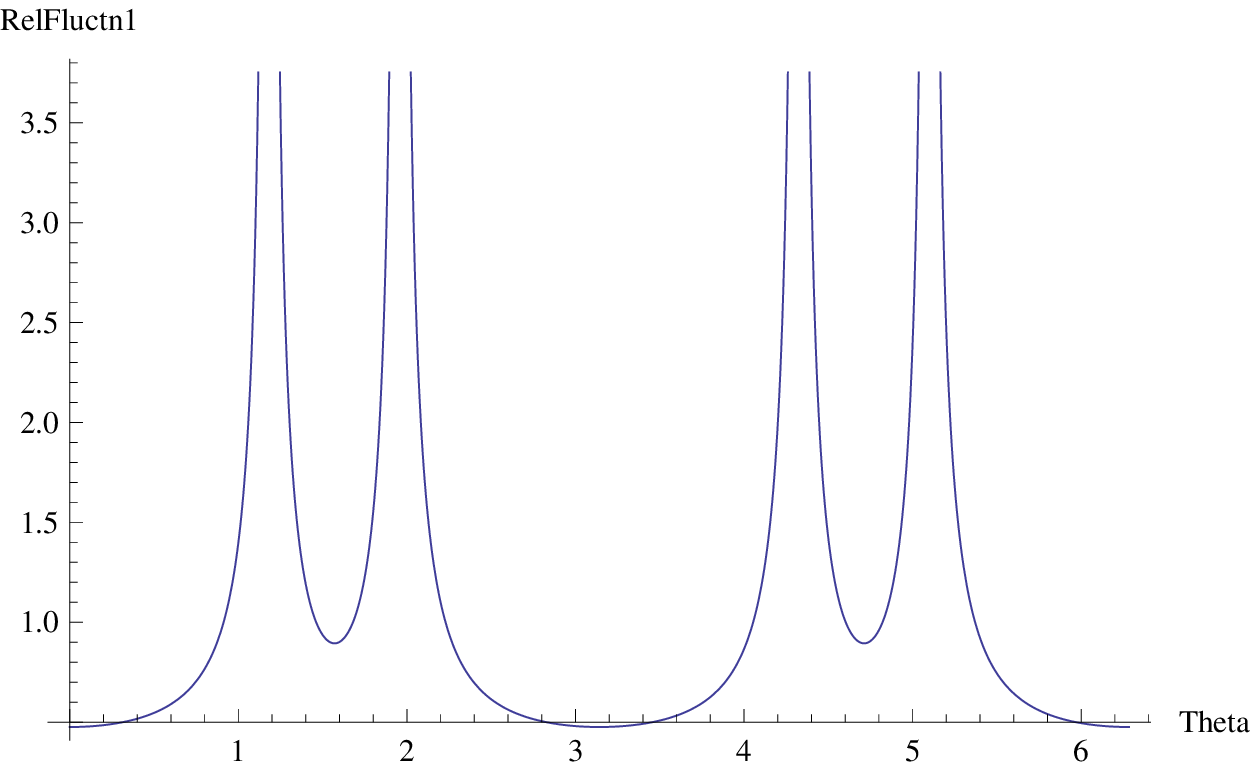}%
\end{center}
\end{figure}

\bigskip

\begin{center}
Figure 1. Relative energy fluctuation for Josephson parameters in $\lambda
_{1}(\theta_{p})$ case.

\bigskip%
\begin{figure}
[ptb]
\begin{center}
\includegraphics[
height=3.0364in,
width=5.0548in
]%
{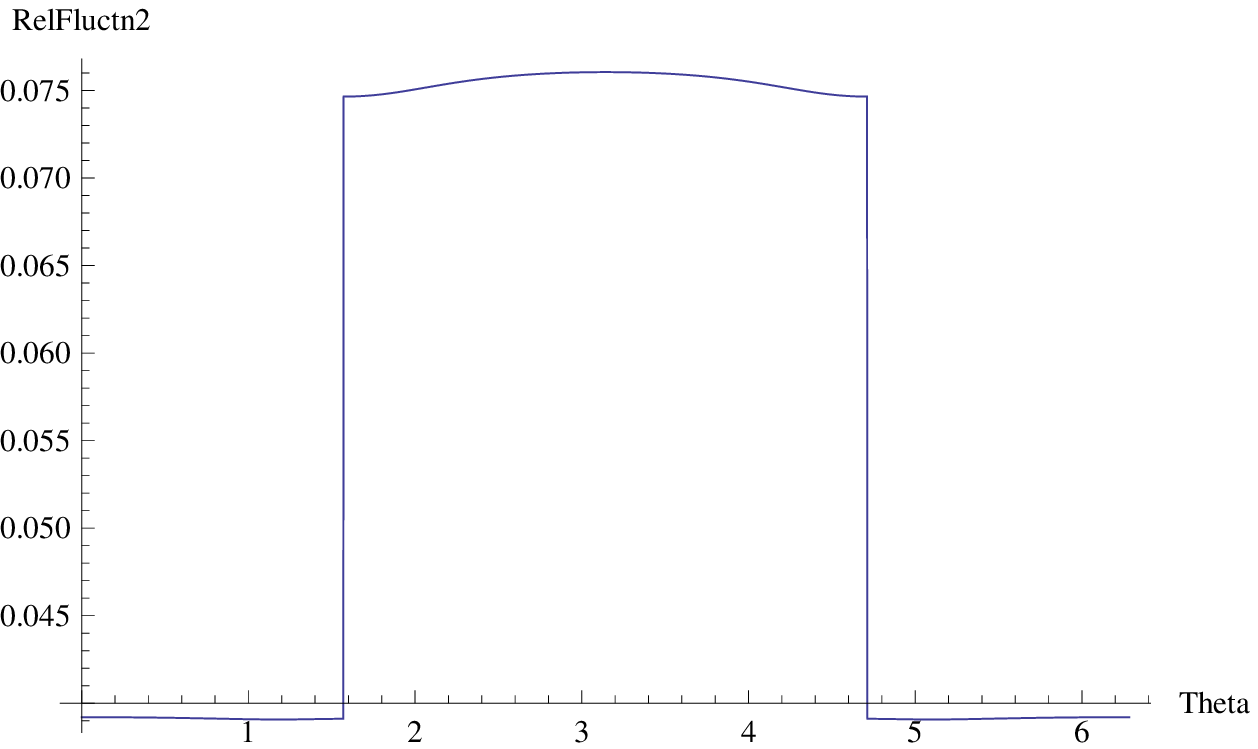}%
\end{center}
\end{figure}

\bigskip

Figure 2. Relative energy fluctuation for Josephson parameters in $\lambda
_{2}(\theta_{p})$ case.\bigskip
\end{center}

Clearly for the choice of Josephson parameters in the $\lambda_{1}(\theta
_{p})$ case the relative energy fluctuations are very large, $O(200\%)$.
However, for the choice of Josephson parameters in the $\lambda_{2}(\theta
_{p})$ case the relative energy fluctuations are fairly small, $O(6\%)$. If
the Josephson parameters are chosen as in the latter case, then an adiabatic
process starting with parameters as for some initial $\theta_{p0}$ and
changing them to those for $\theta_{p}$ in accordance with
Eq.(\ref{Eq.SpecialJosephParam}) with $\alpha=2$ could prepare a state
\textit{close} to the required relative phase state. For example, with
$\theta_{p0}=\frac{{\LARGE \pi}}{{\LARGE 2}}$ we have $UN=0$, $J=K$ and
$\delta=0$ since $X_{12}=0$, $X_{22}=1$ and $X_{32}=0$. A suitable initial
state within the Rabi regime where $J\gg UN$, $\delta=0$ might be used. In
particular, the state discussed below corresponding to that created at the end
of the first stage in the proposed Heisenberg limited interferometry
experiment has a quite well defined relative phase $\theta_{p0}=0$ (see Figure
4 below) and might be suitable. The required quantities $X_{12\text{ }}$ and
$X_{22}$ that define the way $UN$, $J$ would be adiabatically changed to reach
any required $\theta_{p}$ are shown in Figure 3 (formulae are also given in
the Appendix). $\delta$ would remain equal to zero.\medskip%
\begin{figure}
[ptb]
\begin{center}
\includegraphics[
height=2.8954in,
width=5.0548in
]%
{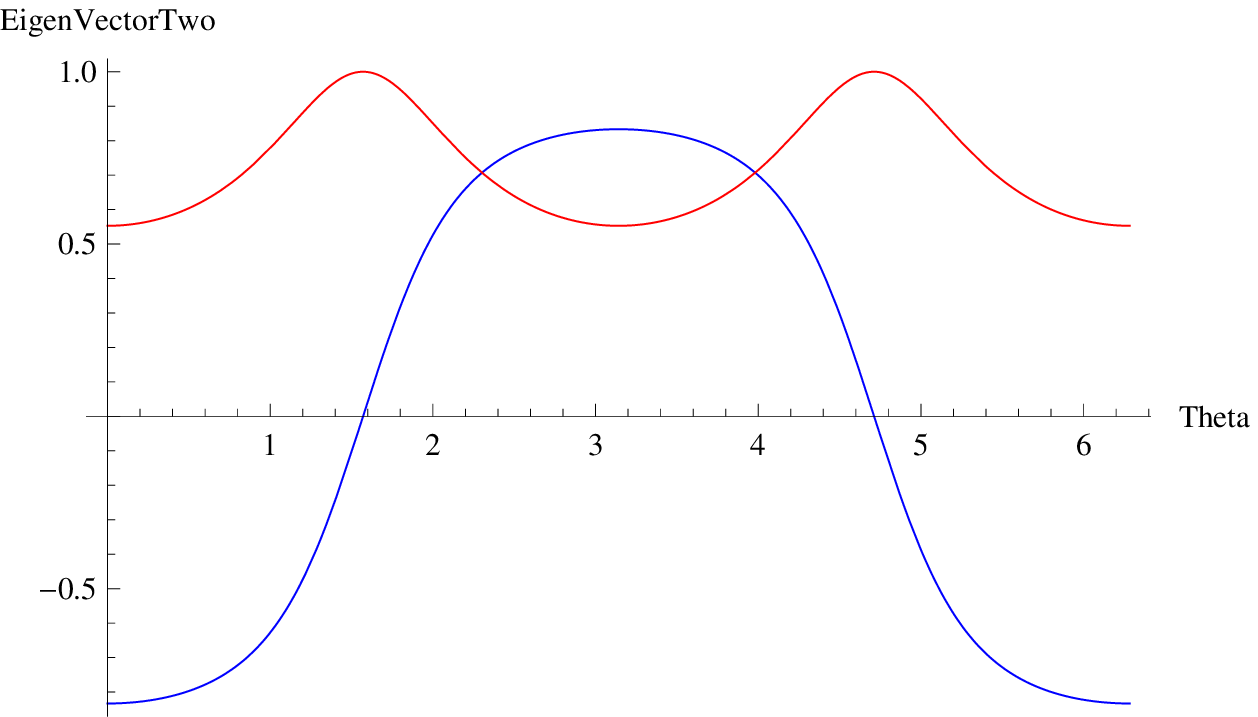}%
\end{center}
\end{figure}

\bigskip

\begin{center}
Figure 3. Eigenvector for $\lambda_{2}(\theta_{p})$. $X_{12}$ (blue curve) and
$X_{22}$ (red curve). $X_{32}=0$.
\end{center}

\bigskip

No actual experiment for preparing a BEC in a relative phase eigenstate or for
directly measuring the relative phase probability have yet been carried out,
so the relative phase probability distribution may need to be inferred from
other measurements rather than directly measured. However, similar remarks may
be made about position eigenstates for individual particles - where only
states with relatively localised positions can be prepared and where position
probabilty results are inferred from experiments involving scattering of weak
probe beams, so this need not preclude the relative phase operator and its
eigenstates being useful concepts in quantum atom optics.\medskip

\subsection{Interferometry experminents an quantum correlation functions}

The quantum correlation function $G^{(1)}(\mathbf{r},\mathbf{r}^{\prime})$
with $\mathbf{r}=\mathbf{r}^{\prime}$ is of particular interest as it
determines the probability distribution for boson position measurements
\cite{Bach04a}, \cite{Bach04b} and hence is useful in describing the
interference fringes that can occur in BEC interferometry experiments. For a
general state (\ref{Eq.PureStatePhaseStatesExpn}) the first order quantum
correlation function can also be expressed in terms of the amplitudes
$A(\theta_{p})$ for the relative phase eigenstates. These amplitudes appear
via three \textit{autocorrelation functions}. We have%
\begin{align}
& G^{(1)}(\mathbf{r},\mathbf{r}^{\prime})\nonumber\\
& =\phi_{a}{\small (\mathbf{r})}^{\ast}\phi_{a}{\small (\mathbf{r}}^{\prime
}{\small )}\sum_{r}C_{0}(\theta_{r})\frac{1}{N+1}%
{\textstyle\sum\limits_{k}}
\exp(ik\theta_{r})\left(  \frac{N}{2}-k\right) \nonumber\\
& +\phi_{b}{\small (\mathbf{r})}^{\ast}\phi_{b}{\small (\mathbf{r}}^{\prime
}{\small )}\sum_{r}C_{0}(\theta_{r})\frac{1}{N+1}%
{\textstyle\sum\limits_{k}}
\exp(ik\theta_{r})\left(  \frac{N}{2}+k\right) \nonumber\\
& +\phi_{a}{\small (\mathbf{r})}^{\ast}\phi_{b}{\small (\mathbf{r}}^{\prime
}{\small )}\sum_{r}C_{+1}(\theta_{r})\frac{1}{N+1}%
{\textstyle\sum\limits_{k}}
\exp(ik\theta_{r})\,\sqrt{\left(  \frac{N}{2}-k+1\right)  \left(  \frac{N}%
{2}+k\right)  }\nonumber\\
& +\phi_{b}{\small (\mathbf{r})}^{\ast}\phi_{a}{\small (\mathbf{r}}^{\prime
}{\small )}\sum_{r}C_{-1}(\theta_{r})\frac{1}{N+1}%
{\textstyle\sum\limits_{k}}
\,\exp(ik\theta_{r})\sqrt{\left(  \frac{N}{2}+k+1\right)  \left(  \frac{N}%
{2}-k\right)  }\label{Eq.G1ResultPhaseStates}%
\end{align}
where the autocorrelation functions of the amplitudes $A(\theta_{p})$ are
defined as%
\begin{align}
C_{0}(\theta_{r})  & =%
{\textstyle\sum\limits_{q=-\frac{\mathbf{N}}{\mathbf{2}}}^{\frac{\mathbf{N}%
}{\mathbf{2}}}}
\,A(\{\theta_{r}+\theta_{q}\}_{\operatorname{mod}2\pi})\,A(\theta_{q})^{\ast
}\nonumber\\
C_{\pm1}(\theta_{r})  & =%
{\textstyle\sum\limits_{q=-\frac{\mathbf{N}}{\mathbf{2}}}^{\frac{\mathbf{N}%
}{\mathbf{2}}}}
\,A(\{\theta_{r}+\theta_{q}\}_{\operatorname{mod}2\pi})\,A(\theta_{q})^{\ast
}\exp(\pm i\theta_{q})\label{Eq.AutoCorrnPhase2}%
\end{align}
Hence we see that for a state with a relative narrow relative phase
distribution around a particular phase $\theta_{0}$ ($A(\theta_{p}%
)\approx\delta_{\theta_{p},\theta_{0}}$) the autocorrelation function
$C_{0}(\theta_{r})$ will be peaked around $\theta_{r}\doteqdot0$, whilst the
autocorrelation functions $C_{\pm1}(\theta_{r})$ will be peaked around
$\theta_{r}\doteqdot\mp\theta_{0}$. This means that the first two terms in
$G^{(1)}(\mathbf{r},\mathbf{r}^{\prime})$ have no dependence on $\theta_{0}$,
whereas the last two terms have essentially a sinusoidal variation with
$\theta_{0}$ since the sum over $\theta_{r}$ will be dominated by terms
$\theta_{r}\doteqdot\mp\theta_{0}$. If the particular central phase
$\theta_{0}$ is changed during an experiment then the boson position
probability will change - hence a fringe pattern would be observed. Note that
the observation of the fringe depends on the overlap of the mode functions
being sufficiently large. On the other, for a state with a relatively wide
relative phase distribution the auto correlation functions will be significant
for a wide range of $\theta_{r}$ so the sum over $\theta_{r}$ will be no
longer be dominated by terms $\theta_{r}\doteqdot\mp\theta_{0}$ and the fringe
pattern would be washed out.\medskip

\subsection{Heisenberg limited BEC interferometry experiment}

The relative phase eigenstate is a valuable theoretical concept for describing
the behaviour in BEC\ interferometry experiments. For example, in the proposed
experiment by Dunningham and Burnett \cite{Dunningham04a} for Heisenberg
limited interferometry in two mode BEC, the collapse and revival of
interference fringes can be discussed in terms of collapses and revivals of
the time dependent probability distribution for the relative phase. Collapse
and revival effects in BEC were described earlier by Wright et al
\cite{Wright96a}.

The Dunningham and Burnett experiment treats the two mode double well BEC
system via the Josephson Hamiltonian. The experiment has two stages. In the
first stage the system starts with equal numbers of bosons in each well, so
the quantum state is $\left\vert \Phi(0)\right\rangle =\left\vert
N/2\right\rangle _{a}\left\vert N/2\right\rangle _{b}$, so $b_{k}%
(0)=\delta_{k,0}$. From Eqs.(\ref{Eq.ProbRelPhase}) and (\ref{Eq.RelnCoefts})
it is easy to see that the relative phase probability distribution is uniform.
With evolution dominated by the tunneling term the state evolves for a time
$T_{1}=\pi\hbar/2J$ (or when $\phi=Jt/\hbar=\pi/2$) \ The methods of angular
momentum theory can be used to determine the dynamics, since the evolution
operator $\widehat{U}(t)=\exp(i\widehat{S}_{x}Jt/\hbar)$ is just a rotation
operator. We find that%
\begin{equation}
b_{k}(T_{1})=\exp(ik\frac{\pi}{2})\frac{\sqrt{(N/2+k)!(N/2-k)!}}%
{(N/2)!(2^{N/2})}%
{\displaystyle\sum\limits_{p}}
(-1)^{p}C_{p+k}^{N/2}C_{p}^{N/2}\label{Eq.AmpT1}%
\end{equation}
from which we can calculate the relative phase probability distribution via
Eqs.(\ref{Eq.ProbRelPhase}) and (\ref{Eq.RelnCoefts}). This is shown in Figure
4 for the case $N=80$ and we see that the system has a well-defined relative
phase of approximately zero. This stage of the experiment involves creating a
state with a rather well-defined relative phase. Experimentally the time
$T_{1}$ is determined by observing the time it takes for the fringe pattern to
become sharpest.

\bigskip\
\begin{figure}
[ptb]
\begin{center}
\includegraphics[
height=3.1324in,
width=5.0548in
]%
{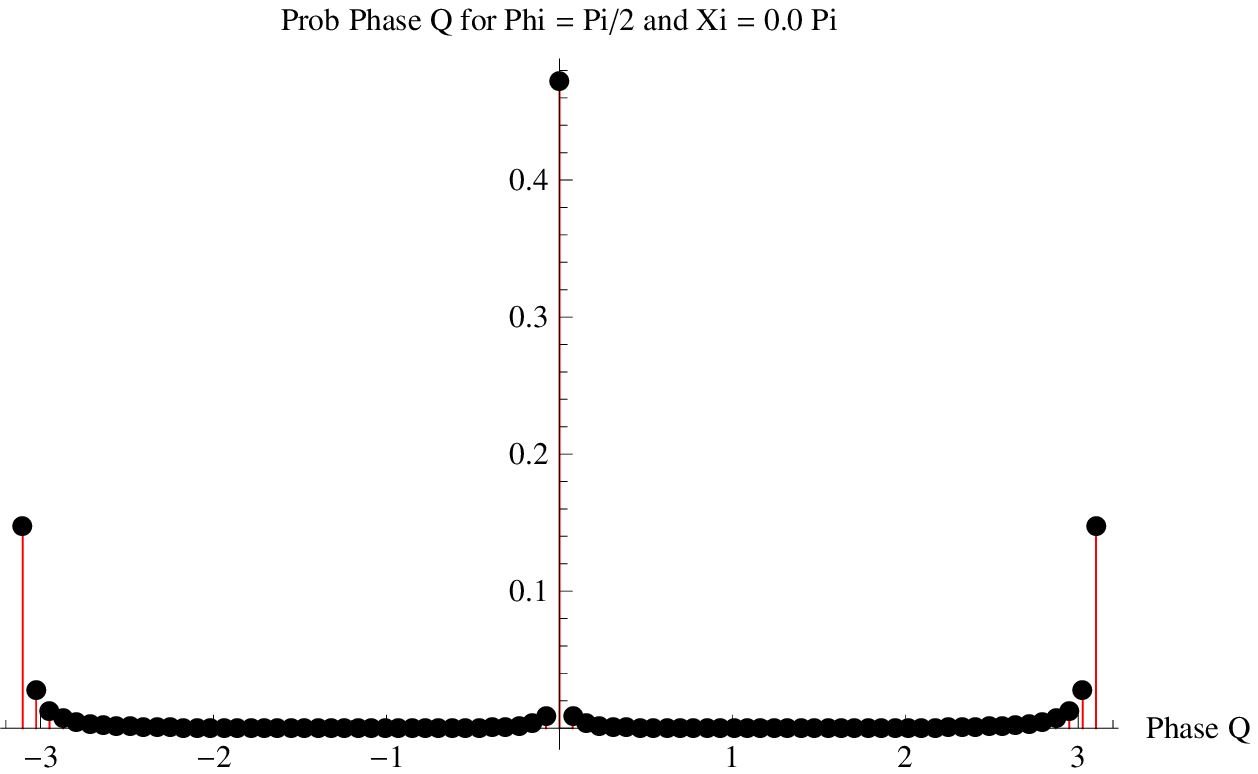}%
\end{center}
\end{figure}

\begin{center}
Figure 4. Relative phase probability at end of the first stage. Well defined
relative phase seen. Parameters are given in text. \bigskip
\end{center}

The second stage involves evolution for a further time $T$ dominated by the
collision term, or the collision term plus the asymmetry term. In this case we
find that%
\[
b_{k}(T_{1}+T)=\exp(-ik\frac{\delta T}{\hbar})\exp(-ik^{2}\frac{UT}{\hbar
})b_{k}(T_{1})
\]
We first consider the situation when there is no asymmetry $\delta=0$. A
characteristic time scale for the collision dominated evolution is $T_{2}%
=\pi\hbar/2U$ (or when $\xi=Ut/\hbar=\pi/2$). However, due to the
$\exp(-ik^{2}\frac{UT}{\hbar})$ factor there is a dephasing effect, causing
the relative phase probability amplitudes $A(\theta_{p},T_{1}+T)$ to become
significant over a wide range of $\theta_{p}$. This causes a collapse in the
previously well defined interference fringe pattern. The time scale for this
to happen is that required for the fastest pairs of contributions
($k=-N/2,-N/2+1$ or $k=+N/2,+N/2-1$) to get out of phase by $\sim\pi$. Thus
the collapse time is given by $T_{c}=\pi\hbar/NU=T_{2}/N$ (or when $\xi
=\pi/2N$), which is $O(1/N)$ times shorter than $T_{2}$. This collapse effect
is shown in Figures 5 and 6 for $N=80$. For $\xi=0.001\pi$ the relative phase
distribution is starting to spread out and is essentially uniform when
$\xi=0.01\pi$.

\bigskip%
\begin{figure}
[ptb]
\begin{center}
\includegraphics[
height=3.1324in,
width=5.0548in
]%
{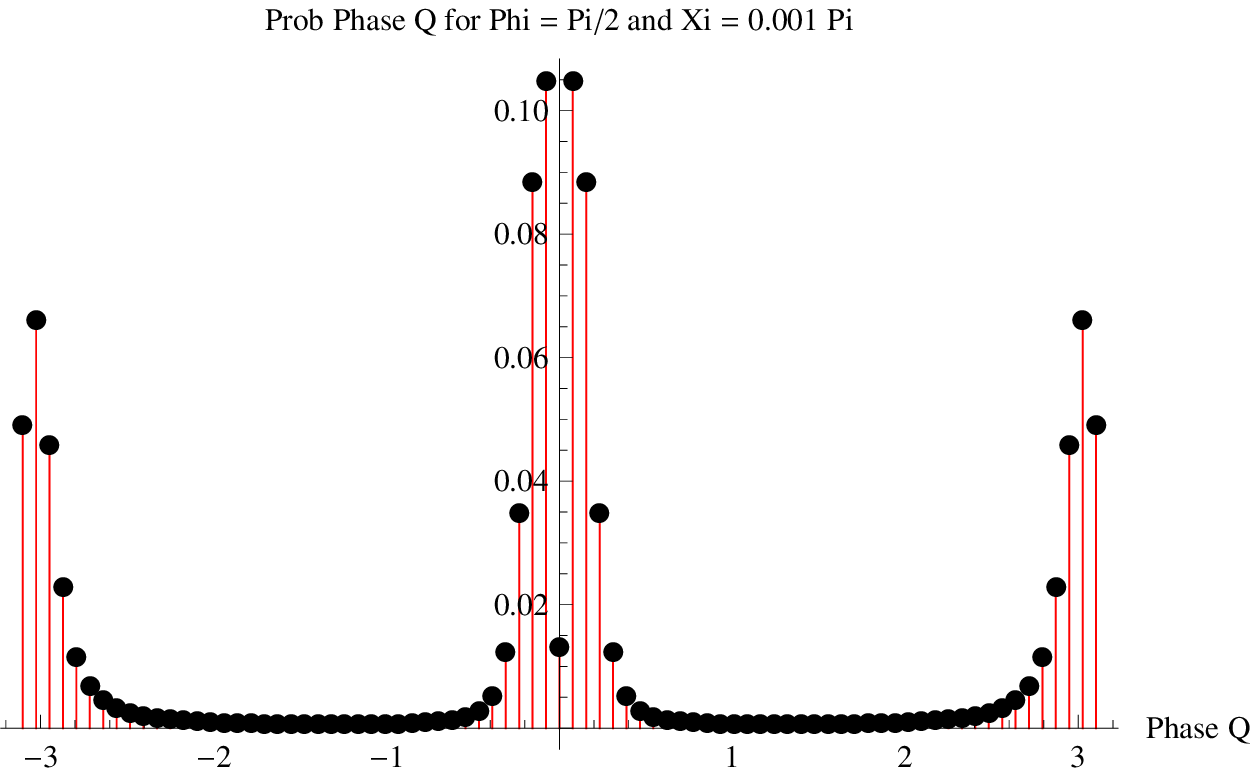}%
\end{center}
\end{figure}
\ 

\begin{center}
Figure 5. Relative phase probability just after the end of the first stage.
Dephasing effects starting to be seen. Parameters are given in text. \bigskip%
\begin{figure}
[ptb]
\begin{center}
\includegraphics[
height=3.1324in,
width=5.0548in
]%
{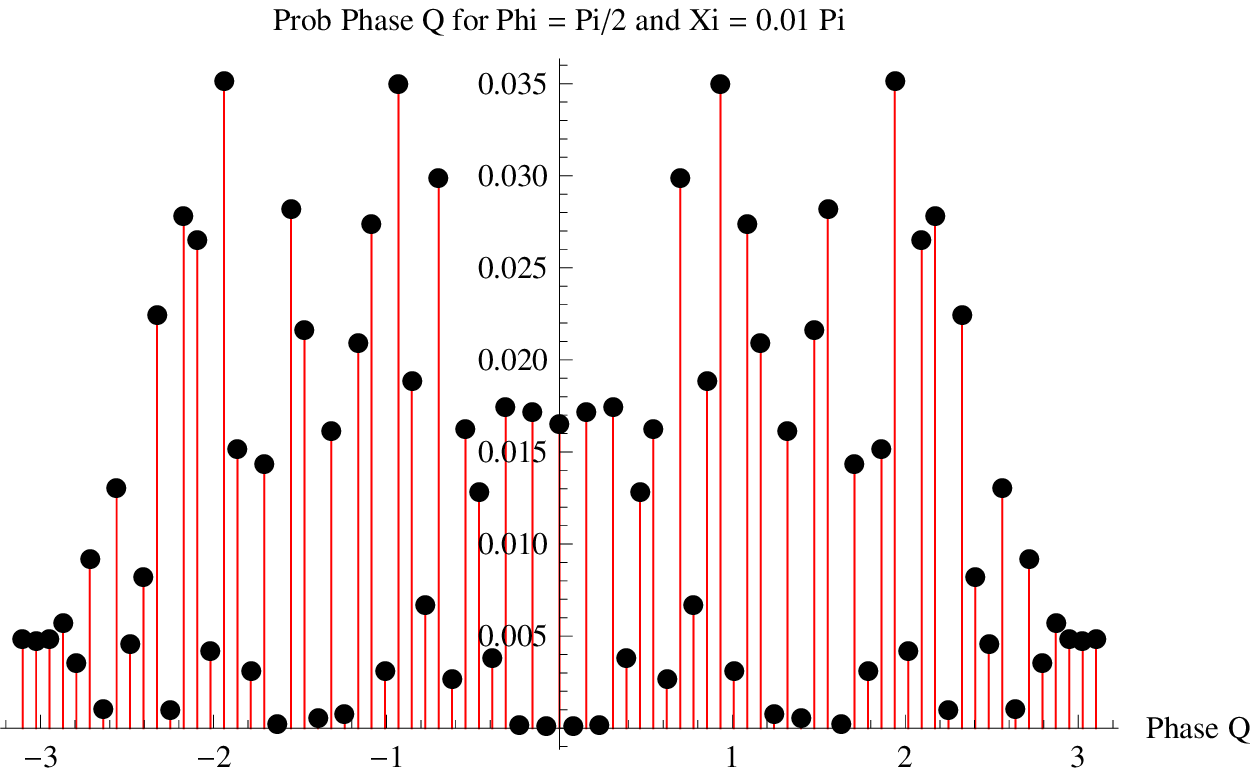}%
\end{center}
\end{figure}

Figure 6. Relative phase probability somewhat after the end of the first
stage. Complete dephasing effects seen. Parameters are given in text. \bigskip
\end{center}

However, the factors $\exp(-ik^{2}\frac{UT}{\hbar})$ do eventually get back
into phase. If $\xi=Ut/\hbar$ is a multiple of $\pi/2$ then all the phase
factors have a modulus of unity, irrespective of $k$. Hence a revival of the
relative phase probability distribution to the sharply defined distribution
that occured at the end of the first stage will take place. The revival time
scale is thus given by $T_{rev}=T_{2}=\pi\hbar/2U$ (or when $\xi=Ut/\hbar
=\pi/2$). This is shown in Figure 7 for $N=80$. In Figure 8 the time is
slightly longer than $T_{rev}$ and the relative phase distribution is starting
to collapse again.\bigskip%
\begin{figure}
[ptb]
\begin{center}
\includegraphics[
height=3.1324in,
width=5.0548in
]%
{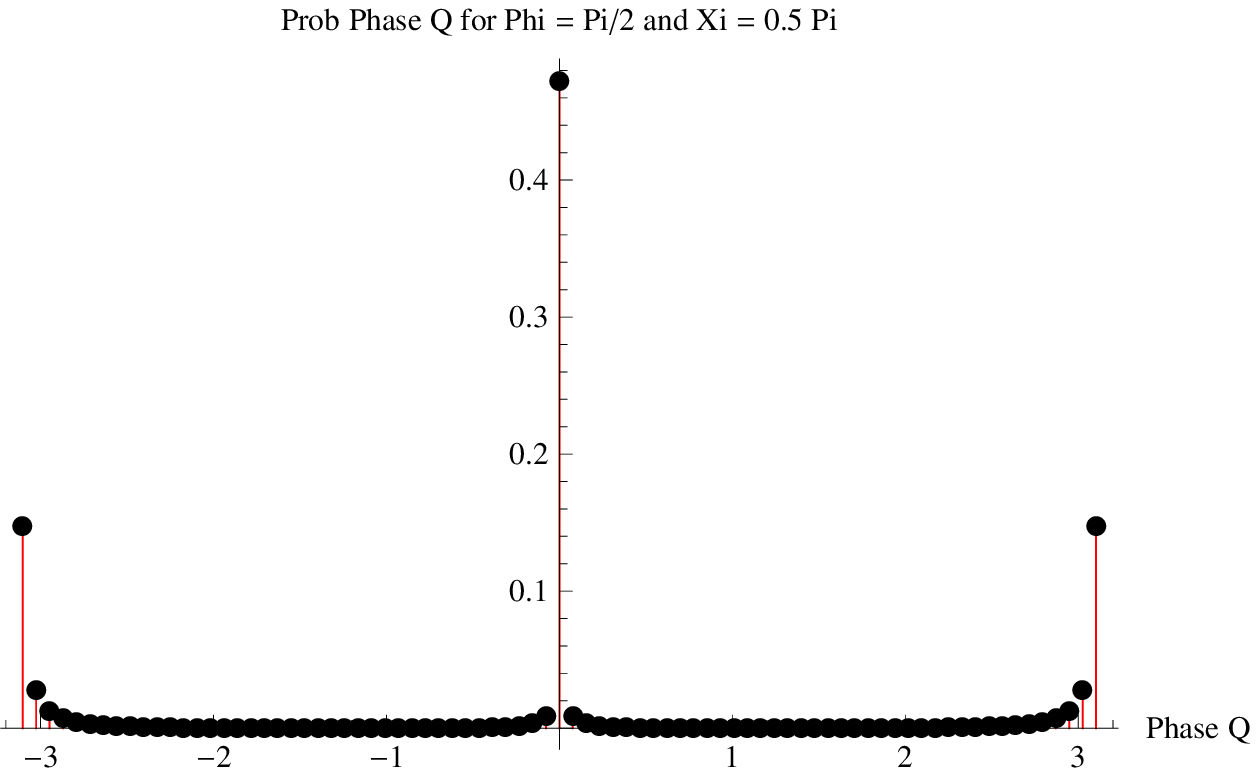}%
\end{center}
\end{figure}

\begin{center}
Figure 7. Relative phase probability at the end of the second stage. Revival
of well-defined phase seen. Parameters are given in text. \bigskip%
\begin{figure}
[ptb]
\begin{center}
\includegraphics[
height=3.1324in,
width=5.0548in
]%
{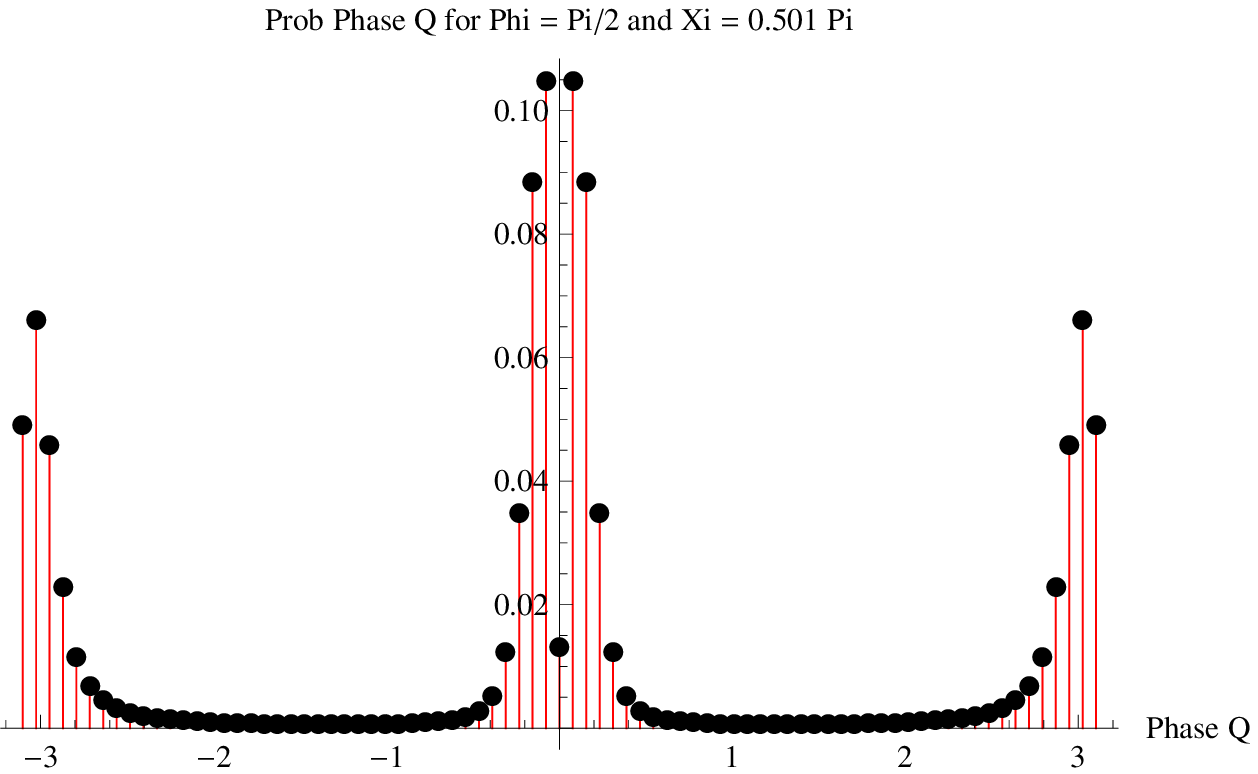}%
\end{center}
\end{figure}

Figure 8. Relative phase probability just after the end of the second stage.
Beginning of collapse of well-defined phase seen. Parameters are given in
text. \bigskip
\end{center}

We now consider the effect of asymmetry. It is easy to see that at time
$T_{1}+T_{2}$ the relative phase amplitude for non zero $\delta$ is given by%
\[
A(\theta_{p},T_{1}+T_{2})=A(\theta_{p}+\delta T_{2}/\hbar,T_{1}+T_{2}%
)_{\delta=0}
\]
so is of the same form as when there is no asymmetry, but with the relative
phase shifted by $\delta T_{2}/\hbar=(\delta/U)\pi/2$. This effect is shown in
Figure 9 for $N=80$. The shift in the fringe pattern would be observable if
$\delta$ is a reasonable fraction of $U$.\bigskip%
\begin{figure}
[ptb]
\begin{center}
\includegraphics[
height=3.1324in,
width=5.0548in
]%
{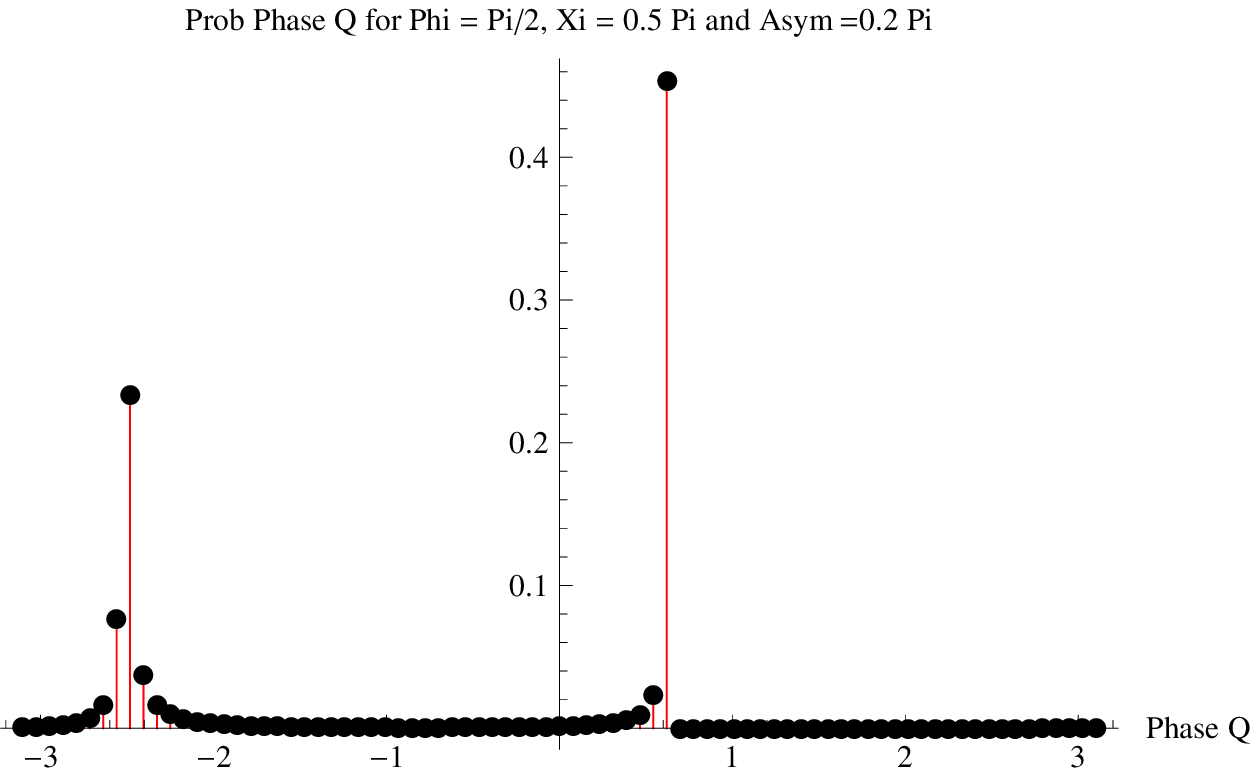}%
\end{center}
\end{figure}

\begin{center}
Figure 9. Relative phase probability at the end of the second stage. Asymmetry
present. Well-defined but shifted phase seen. Parameters are given in text.
\bigskip
\end{center}

Evolution during the second stage of the experiment is allowed to occur for a
time $T_{2}$ corresponding to the revival time, and with zero asymmetry
present. The revival time could be determined experimentally by observing when
the sharp fringes obtained at the end of the first stage are restored again.
The accuracy in determining the revival tiime is given by the collapse time,
so the fractional error in the revival time $T_{rev}$ is of order $1/N$. If
the second stage is run again with asymmetry present the fringe pattern at the
revival time is shifted by $\delta T_{rev}/\hbar$. If this phase shift is
measured with perfect accuracy, then the fractional error in measuring
$\delta$ is the same as that for $T_{rev}$, and hence is of order $1/N$. This
represents a Heisenberg limited interferometry measurement of the asymmetry,
scaling as the inverse of the total number of bosons.\medskip

\section{Summary}

This paper presents an approach to treating phase in quantum atom optics in
which phase is regarded as a physical quantity for the system and treated
theoretically as a linear Hermitian operator, similar to the Pegg-Barnett
treatment of phase in quantum optics. Other approaches to treating phase are
discussed and a brief review outlines previous attempts to find a Hermitian
phase operator. The Pegg-Barnett approach is adapted to define a relative
phase operator for two mode systems via first introducung a complete
orthonormal set of relative phase eigenstates. These states are contrasted
with other so-called phase states. The entanglement, fragmentation and spin
squeezing properties of the relative phase states are set out. The relative
phase state has maximal two mode entanglement, it is a fragmented state with
its Bloch vector lying inside the Bloch sphere and is highly spin squeezed. In
the final section applications of the relative phase states in describing BEC
interferometry experiments are made, both in general and in the context of a
proposed Heisenberg limited interferometry experiment. Interferometry
experiments essentially measure the autocorrelation functions for the relative
phase amplitudes. The possibility for preparing a BEC in a relative phase
state is examined, and an approach based on adiabatically changing parameters
in the Josephson Hamiltonian is suggested. However, the difficulty is similar
to preparing a particle system in a position eigenstate. In spite of this, the
relative phase states are still a useful concept for describing experments in
quantum atom optics. Finally, if such a highly spin squeezed state could be
prepared it may be useful for Heisenberg limited interferometry in view of the
fractional fluctuation in one of the spin operator components perpendicular to
the Bloch vector being essentially only of order $1/N$. \medskip

\section{Acknowledgements}

This work was supported by the Australian Research Council Centre of
Excellence for Quantum Atom Optics. The author thanks F. Baumgartner, S.
Barnett, T. Busch, J. Close, J. Dingjan, P. Drummond, M. Egorov, E. Hinds and
A. Sidorov for helpful discussions. 

\begin{center}
\pagebreak
\end{center}

\section{Appendix}

In the appendix we summarize certain key results needed in the main body of
the paper.

\subsection{Field operators}

For the two mode BEC the field operators are given in the two mode
approximation as
\begin{equation}
\widehat{\Psi}(\mathbf{r})=\widehat{a}\phi_{a}(\mathbf{r})+\widehat{b}\phi
_{b}(\mathbf{r})\qquad\widehat{\Psi}(\mathbf{r})^{\dag}=\widehat{a}^{\dag}%
\phi_{a}^{\ast}(\mathbf{r})+\widehat{b}^{\dag}\phi_{b}^{\ast}(\mathbf{r}%
)\label{Eq.FieldOprs}%
\end{equation}
and with the usual non-zero commutation rules for the mode operators
$[\widehat{a},\widehat{a}^{\dag}]=[\widehat{b},\widehat{b}^{\dag}]=\widehat
{1},$the non-zero commutation rule for the field operators is
\begin{equation}
\lbrack\widehat{\Psi}(\mathbf{r}),\widehat{\Psi}(\mathbf{r}^{\prime})^{\dag
}]=\phi_{a}(\mathbf{r})\phi_{a}^{\ast}(\mathbf{r}^{\prime})+\phi
_{b}(\mathbf{r})\phi_{b}^{\ast}(\mathbf{r}^{\prime})=\delta_{2}(\mathbf{r,r}%
^{\prime})\label{Eq.FieldOprCommRule}%
\end{equation}
which is a restricted delta function for the space spanned by the two
orthonormal mode functions.\smallskip

\subsection{Spin operators and spin states}

Because of the two-mode approximation it is possible to treat the bosonic
system using the methods of \textit{angular momentum theory}. The system
behaves like a macroscopic spin system with angular momentum quantum number
$j=\frac{{\LARGE N}}{{\LARGE 2}}$.

In a two-mode theory it is convenient to introduce the Schwinger \textit{spin
angular momentum operators} defined by%
\begin{align}
\widehat{S}_{x}  & =(\widehat{b}^{\dag}\widehat{a}+\widehat{a}^{\dag}%
\widehat{b})/2\nonumber\\
\widehat{S}_{y}  & =(\widehat{b}^{\dag}\widehat{a}-\widehat{a}^{\dag}%
\widehat{b})/2i\nonumber\\
\widehat{S}_{z}  & =(\widehat{b}^{\dag}\widehat{b}-\widehat{a}^{\dag}%
\widehat{a})/2\label{Eq.AngMtmOprs}%
\end{align}
The spin operators $\widehat{S}_{a}$ satisfy the standard commutation rules
for angular momentum operators%

\begin{equation}
\left[  \widehat{S}_{a}\mathbf{,}\widehat{S}_{b}\right]  =i\,\epsilon
_{abc}\widehat{S}_{c}\quad\quad{\small (}a,b,c=x,y,z{\small ),}%
\label{Eq.AngularMtmCommRules}%
\end{equation}
and the \textit{square} of the angular momentum $(\underrightarrow{\widehat
{S}})^{2}$ can be related to the boson number operator. Thus:
\begin{align}
(\underrightarrow{\widehat{S}})^{2}  & =%
{\textstyle\sum\limits_{a}}
(\widehat{S}_{a})^{2}\nonumber\\
& =\frac{\widehat{N}}{2}(\frac{\widehat{N}}{2}+1)\label{Eq.AngMtmSquare}%
\end{align}
where
\begin{equation}
\widehat{N}=(\widehat{b}^{\dag}\widehat{b}+\widehat{a}^{\dag}\widehat
{a})\label{Eq.NumberOpr}%
\end{equation}
is the number operator. Clearly the angular momentum squared is a
\textit{conserved} quantity. Note that the spin operator $\widehat{S}_{z}$ is
the same as the relative number operator $\delta\widehat{N}$.

The $N$ boson system behaves like a \textit{giant spin system} in the two-mode
approximation. The basis states $\left\vert N/2-k\right\rangle _{a}\left\vert
N/2+k\right\rangle _{b}$ are simultaneous eigenstates of $(\underrightarrow
{\widehat{S}})^{2}$ and $\widehat{S}_{z}$ with eigenvalues $\frac{{\LARGE N}%
}{{\LARGE 2}}(\frac{{\LARGE N}}{{\LARGE 2}}+1)$ and $k$ respectively. To
emphasize the spin character of the basis states we can introduce the notation%
\begin{align}
\left\vert N/2-k\right\rangle _{a}\left\vert N/2+k\right\rangle _{b}  &
\equiv\frac{\left(  \widehat{a}^{\dag}\right)  ^{(\frac{{\LARGE N}}%
{{\LARGE 2}}-k)}}{[(\frac{N}{2}-k)!]^{\frac{\mathbf{1}}{\mathbf{2}}}}%
\frac{\left(  \widehat{b}^{\dag}\right)  ^{(\frac{{\LARGE N}}{{\LARGE 2}}+k)}%
}{[(\frac{N}{2}+k)!]^{\frac{\mathbf{1}}{\mathbf{2}}}}\left\vert
\,0\right\rangle \nonumber\\
& \equiv\left\vert \,\frac{{\small N}}{{\small 2}},k\right\rangle
\label{Eq.BasisStateNotation}%
\end{align}
Thus:
\begin{align}
(\underrightarrow{\widehat{S}})^{2}\,\left\vert \,\frac{{\small N}}%
{{\small 2}},k\right\rangle  & =\frac{{\small N}}{{\small 2}}(\frac
{{\small N}}{{\small 2}}+{\small 1})\,\left\vert \,\frac{{\small N}%
}{{\small 2}},k\right\rangle \label{Eq.SpinSquaredEigenEqn}\\
\widehat{S}_{z}\,\left\vert \,\frac{{\small N}}{{\small 2}},k\right\rangle  &
=k\,\left\vert \,\frac{{\small N}}{{\small 2}},k\right\rangle
\label{Eq.SpinZEigenEqn}%
\end{align}
Hence $j=\frac{N}{2}$ is the \textit{spin angular momentum} quantum number,
and $k$ is the \textit{spin magnetic} quantum number, with $(-\frac{N}{2}\leq
k\leq\frac{N}{2})$. Thus the boson number $N$ and the quantity $k$ that
specifies the fragmentation of the BEC between the two modes have a physical
interpretation in terms of angular momentum theory. Since boson numbers may be
$\sim10^{8}$ the spin system is on a macroscopic scale. \smallskip

As in angular momentum theory we find it convenient to introduce \textit{spin
up} $\widehat{S}_{+}$ and \textit{spin down} $\widehat{S}_{-}$ operators,
which change the spin magnetic quantum numbers by $\pm1$. We have
\begin{align}
\widehat{S}_{\pm}\,\left\vert \,\frac{{\small N}}{{\small 2}},k\right\rangle
& =\{\frac{{\small N}}{{\small 2}}(\frac{{\small N}}{{\small 2}}%
+{\small 1})-k(k\pm{\small 1})\}^{\frac{\mathbf{1}}{\mathbf{2}}}\,\left\vert
\,\frac{{\small N}}{{\small 2}},k\pm{\small 1}\right\rangle \nonumber\\
& =\{(\frac{{\small N}}{{\small 2}}\mp k)(\frac{{\small N}}{{\small 2}}\pm
k+{\small 1})\}^{\frac{\mathbf{1}}{\mathbf{2}}}\,\left\vert \,\frac
{{\small N}}{{\small 2}},k\pm{\small 1}\right\rangle \nonumber\\
\widehat{S}_{\pm}  & =\widehat{S}_{x}\pm i\widehat{S}_{y}%
.\label{Eq.SpinUpDownOprs}%
\end{align}

The methods of angular momentum theory can be utilized by first writing the
full Hamiltonian in terms of spin operators using equations
(\ref{Eq.FieldOprs}), (\ref{Eq.AngMtmOprs}), (\ref{Eq.NumberOpr}) - noting
that all terms involve equal numbers of creation and annihilation operators,
and its matrix elements calculated using angular momentum theory from
(\ref{Eq.SpinZEigenEqn}) and (\ref{Eq.SpinUpDownOprs}). The same applies in a
simplification to the full Hamiltonian giving rise to the Josephson
Hamiltonian.\smallskip

\subsection{Quantum correlation functions}

The first order quantum correlation function is defined as
\begin{equation}
G^{(1)}(\mathbf{r,r}^{\prime})=\left\langle \widehat{\Psi}(\mathbf{r})^{\dag
}\widehat{\Psi}(\mathbf{r}^{\prime})\right\rangle \label{Eq.QuantCorrFn}%
\end{equation}
and for the pure state given by (\ref{Eq.PureStateNumberStatesExpn}) this is
found to be%
\begin{align}
G^{(1)}({\small \mathbf{r}}\mathbf{;r}^{\prime}{\small \mathbf{)}}  & =%
{\textstyle\sum\limits_{k}}
\,b_{k}{}^{\ast}b_{k}\left\{  \phi_{a}{\small (\mathbf{r})}^{\ast}\phi
_{a}{\small (\mathbf{r}}^{\prime}{\small )}\left(  \frac{N}{2}-k\right)
+\phi_{b}{\small (\mathbf{r})}^{\ast}\phi_{b}{\small (\mathbf{r}}^{\prime
}{\small )}\left(  \frac{N}{2}+k\right)  \right\} \nonumber\\
& +%
{\textstyle\sum\limits_{k}}
\,b_{k}{}^{\ast}b_{k+1}\left\{  \phi_{a}{\small (\mathbf{r})}^{\ast}\phi
_{b}{\small (\mathbf{r}}^{\prime}{\small )}\sqrt{\left(  \frac{N}{2}-k\right)
\left(  \frac{N}{2}+k+1\right)  }\right\} \nonumber\\
& +%
{\textstyle\sum\limits_{k}}
\,b_{k}{}^{\ast}b_{k-1}\left\{  \phi_{b}{\small (\mathbf{r})}^{\ast}\phi
_{a}{\small (\mathbf{r}}^{\prime}{\small )}\sqrt{\left(  \frac{N}{2}+k\right)
\left(  \frac{N}{2}-k+1\right)  }\right\} \label{Eq.G1Result}%
\end{align}

\subsection{Bloch vector}

The components $S_{a}$ of the \textit{Bloch vector} are given by averages of
the \textit{spin operators} $\widehat{S}_{a}$
\begin{equation}
S_{a}=\left\langle \widehat{S}_{a}\right\rangle \qquad
(a=x,y,z)\label{Eq.BlochVectorCompts}%
\end{equation}
Often the Bloch vector\emph{\ }components are scaled in units of $N$, but to
avoid extra notation we will not do that here.

Since $\widehat{S}_{a}$ is hermitian and $\left\langle \widehat{S}%
_{a}\right\rangle ^{2}\leq\left\langle (\widehat{S}_{a})^{2}\right\rangle $
and using (\ref{Eq.AngMtmSquare}) we see that
\begin{equation}
0\leq%
{\displaystyle\sum\limits_{a}}
S_{a}^{2}\leq%
{\displaystyle\sum\limits_{a}}
\left\langle (\widehat{S}_{a})^{2}\right\rangle =\frac{N}{2}(\frac{N}%
{2}+1)\doteqdot\frac{N^{2}}{4}\qquad N\gg1\label{Eq.BlochVectorMagnitude0}%
\end{equation}
showing that for all states the Bloch vector lies inside or on a \textit{Bloch
sphere}, whose radius is $\frac{{\LARGE N}}{{\LARGE 2}}$.

For the quantum state given by (\ref{Eq.PureStateNumberStatesExpn})
expressions for the Bloch vector components are%
\begin{align}
S_{\pm}  & =\sum_{k}b_{k}^{\ast}b_{k\mp1}\sqrt{(\frac{{\small N}}{{\small 2}%
}(\frac{{\small N}}{{\small 2}}+{\small 1})-k(k\mp{\small 1})}\nonumber\\
S_{x}  & =(S_{+}+S_{-})/2\qquad S_{y}=(S_{+}-S_{-})/2i\nonumber\\
S_{z}  & =\sum_{k}b_{k}^{\ast}b_{k}\,k\label{Eq.BlochVectorResults}%
\end{align}
and these are related to the first order quantum correlation function via
\begin{align}
& G^{(1)}({\small \mathbf{r}}\mathbf{;r}^{\prime}{\small \mathbf{)}%
}\nonumber\\
& =\left(
\begin{array}
[c]{c}%
\frac{N}{2}\left\{  \phi_{a}{\small (\mathbf{r})}^{\ast}\phi_{a}%
{\small (\mathbf{r}}^{\prime}{\small )}+\phi_{b}{\small (\mathbf{r})}^{\ast
}\phi_{b}{\small (\mathbf{r}}^{\prime}{\small )}\right\} \\
+S_{z}\left\{  -\phi_{a}{\small (\mathbf{r})}^{\ast}\phi_{a}%
{\small (\mathbf{r}}^{\prime}{\small )}+\phi_{b}{\small (\mathbf{r})}^{\ast
}\phi_{b}{\small (\mathbf{r}}^{\prime}{\small )}\right\} \\
+S_{-}\left\{  \phi_{a}{\small (\mathbf{r})}^{\ast}\phi_{b}{\small (\mathbf{r}%
}^{\prime}{\small )}\right\}  +S_{+}\left\{  \phi_{b}{\small (\mathbf{r}%
)}^{\ast}\phi_{a}{\small (\mathbf{r}}^{\prime}{\small )}\right\}
\end{array}
\right)
\end{align}

\subsection{Covariance matrix for spin operators}

The \textit{covariance matri}\emph{x} $C(\widehat{S}_{a},\widehat{S}_{b})$ for
the spin operators $\widehat{S}_{a}$ is given by%

\begin{align}
C(\widehat{S}_{a},\widehat{S}_{b})  & =\frac{1}{2}\left(  \left\langle
\Delta\widehat{S}_{a}\,\Delta\widehat{S}_{b}\right\rangle +\left\langle
\Delta\widehat{S}_{b}\,\Delta\widehat{S}_{a}\right\rangle \right)
\label{Eq.CovarianceMatrix0}\\
\Delta\widehat{S}_{a}  & =\widehat{S}_{a}-\left\langle \widehat{S}%
_{a}\right\rangle \qquad\qquad(a,b=x,y,z)\label{Eq.SpinFluctnOpr}%
\end{align}
where $\Delta\widehat{S}_{a}$ is a spin fluctuation operator. It is easy to
see that the $3\times3$ covariance matrix is real and symmetric and that
$C(\widehat{S}_{a},\widehat{S}_{a})$ gives the \textit{variance} $\left\langle
(\Delta\widehat{S}_{a})^{2}\right\rangle $ for $\widehat{S}_{a}$. These are
the square of the \textit{standard deviations} or \textit{fluctuations}. Such
a matrix defines a positive quadratic form $F(\xi_{x},\xi_{y},\xi_{z})$. With
real $\xi_{a}$ we have
\begin{align}
F(\xi_{x},\xi_{y},\xi_{z})  & =%
{\displaystyle\sum\limits_{a,b}}
\xi_{a}\,C(\widehat{S}_{a},\widehat{S}_{b})\,\xi_{b}\nonumber\\
& =\frac{1}{2}\left(  \left\langle
{\displaystyle\sum\limits_{a}}
\xi_{a}\Delta\widehat{S}_{a}\,%
{\displaystyle\sum\limits_{b}}
\xi_{b}\Delta\widehat{S}_{b}\right\rangle +\left\langle
{\displaystyle\sum\limits_{b}}
\xi_{b}\Delta\widehat{S}_{b}\,%
{\displaystyle\sum\limits_{a}}
\xi_{a}\Delta\widehat{S}_{a}\right\rangle \right) \nonumber\\
& =\left\langle S(\xi)^{\dag}S(\xi)\right\rangle \geq0\label{Eq.QuadraticForm}%
\end{align}
for any state, where $S(\xi)=%
{\displaystyle\sum\limits_{a}}
\xi_{a}\Delta\widehat{S}_{a}=S(\xi)^{\dag}$. Hence the three eigenvalues for
the covariance matrix will be real and positive. Linear combinations of the
$\Delta\widehat{S}_{a} $ involving a real orthogonal matrix will diagonalise
the covariance matrix and the diagonal elements will give the variances for
fluctuations in three orthogonal directions. These specify
the\textit{\ principal quantum fluctuations}.

The covariance matrix can also be written as
\begin{equation}
C(\widehat{S}_{a},\widehat{S}_{b})=\frac{1}{2}\left\langle \left(  \widehat
{S}_{a}\,\widehat{S}_{b}+\widehat{S}_{b}\,\widehat{S}_{a}\right)
\right\rangle -\left\langle \widehat{S}_{a}\right\rangle \left\langle
\widehat{S}_{b}\right\rangle \label{Eq.CovMatrixAntiComm}%
\end{equation}
so it measures the difference between the average of half the
\textit{anti-commutator} of $\widehat{S}_{a},\widehat{S}_{b}$ and the product
of the averages of the separate $\widehat{S}_{a},\widehat{S}_{b}$.\medskip

Expressions for the \textit{covariance matrix elements} for the spin operators
in the case of the pure state given by (\ref{Eq.PureStateNumberStatesExpn})
are as follows.%
\begin{align}
& C_{xx}\nonumber\\
& =C(\widehat{S}_{x},\widehat{S}_{x})\nonumber\\
& =\frac{1}{4}%
{\displaystyle\sum\limits_{k}}
b_{k+2}^{\ast}b_{k}\sqrt{\frac{{\small N}}{{\small 2}}(\frac{{\small N}%
}{{\small 2}}+1)-k(k+1)}\sqrt{\frac{{\small N}}{{\small 2}}(\frac{{\small N}%
}{{\small 2}}+1)-(k+1)(k+2)}\nonumber\\
& +\frac{1}{2}%
{\displaystyle\sum\limits_{k}}
b_{k}^{\ast}b_{k}(\frac{{\small N}}{{\small 2}}(\frac{{\small N}}{{\small 2}%
}+1)-k^{2})\nonumber\\
& +\frac{1}{4}%
{\displaystyle\sum\limits_{k}}
b_{k-2}^{\ast}b_{k}\sqrt{\frac{{\small N}}{{\small 2}}(\frac{{\small N}%
}{{\small 2}}+1)-k(k-1)}\sqrt{\frac{{\small N}}{{\small 2}}(\frac{{\small N}%
}{{\small 2}}+1)-(k-1)(k-2)}\nonumber\\
& -\frac{1}{4}\left(
\begin{array}
[c]{c}%
{\displaystyle\sum\limits_{k}}
b_{k+1}^{\ast}b_{k}\sqrt{\frac{{\small N}}{{\small 2}}(\frac{{\small N}%
}{{\small 2}}+1)-k(k+1)}\\
-%
{\displaystyle\sum\limits_{k}}
b_{k-1}^{\ast}b_{k}\sqrt{\frac{{\small N}}{{\small 2}}(\frac{{\small N}%
}{{\small 2}}+1)-k(k-1)}%
\end{array}
\right)  ^{2}\label{Eq.CovXX}%
\end{align}

and
\begin{align}
& C_{xy}\nonumber\\
& =C(\widehat{S}_{x},\widehat{S}_{y})=C_{yx}\nonumber\\
& =\frac{1}{4i}%
{\displaystyle\sum\limits_{k}}
b_{k+2}^{\ast}b_{k}\sqrt{\frac{{\small N}}{{\small 2}}(\frac{{\small N}%
}{{\small 2}}+1)-k(k+1)}\sqrt{\frac{{\small N}}{{\small 2}}(\frac{{\small N}%
}{{\small 2}}+1)-(k+1)(k+2)}\nonumber\\
& -\frac{1}{4i}%
{\displaystyle\sum\limits_{k}}
b_{k-2}^{\ast}b_{k}\sqrt{\frac{{\small N}}{{\small 2}}(\frac{{\small N}%
}{{\small 2}}+1)-k(k-1)}\sqrt{\frac{{\small N}}{{\small 2}}(\frac{{\small N}%
}{{\small 2}}+1)-(k-1)(k-2)}\nonumber\\
& -\frac{1}{4i}\left(
{\displaystyle\sum\limits_{k}}
b_{k+1}^{\ast}b_{k}\sqrt{\frac{{\small N}}{{\small 2}}(\frac{{\small N}%
}{{\small 2}}+1)-k(k+1)}\right)  ^{2}\label{Eq.CovXY}\\
& +\frac{1}{4i}\left(
{\displaystyle\sum\limits_{k}}
b_{k-1}^{\ast}b_{k}\sqrt{\frac{{\small N}}{{\small 2}}(\frac{{\small N}%
}{{\small 2}}+1)-k(k-1)}\right)  ^{2}\nonumber
\end{align}
and
\begin{align}
& C_{xz}\nonumber\\
& =C(\widehat{S}_{x},\widehat{S}_{z})=C_{zx}\nonumber\\
& =\frac{1}{4}%
{\displaystyle\sum\limits_{k}}
b_{k+1}^{\ast}b_{k}(2k+1)\sqrt{\frac{{\small N}}{{\small 2}}(\frac{{\small N}%
}{{\small 2}}+1)-k(k+1)}\nonumber\\
& +\frac{1}{4}%
{\displaystyle\sum\limits_{k}}
b_{k-1}^{\ast}b_{k}(2k-1)\sqrt{\frac{{\small N}}{{\small 2}}(\frac{{\small N}%
}{{\small 2}}+1)-k(k-1)}\nonumber\\
& -\frac{1}{2}%
{\displaystyle\sum\limits_{k}}
b_{k+1}^{\ast}b_{k}\sqrt{\frac{{\small N}}{{\small 2}}(\frac{{\small N}%
}{{\small 2}}+1)-k(k+1)}%
{\displaystyle\sum\limits_{k}}
b_{k}^{\ast}b_{k}\,k\label{Eq.CovXZ}\\
& -\frac{1}{2}%
{\displaystyle\sum\limits_{k}}
b_{k-1}^{\ast}b_{k}\sqrt{\frac{{\small N}}{{\small 2}}(\frac{{\small N}%
}{{\small 2}}+1)-k(k-1)}%
{\displaystyle\sum\limits_{k}}
b_{k}^{\ast}b_{k}\,k\nonumber
\end{align}
and%
\begin{align}
& C_{yy}\nonumber\\
& =C(\widehat{S}_{y},\widehat{S}_{y})\nonumber\\
& =-\frac{1}{4}%
{\displaystyle\sum\limits_{k}}
b_{k+2}^{\ast}b_{k}\sqrt{\frac{{\small N}}{{\small 2}}(\frac{{\small N}%
}{{\small 2}}+1)-k(k+1)}\sqrt{\frac{{\small N}}{{\small 2}}(\frac{{\small N}%
}{{\small 2}}+1)-(k+1)(k+2)}\nonumber\\
& +\frac{1}{2}%
{\displaystyle\sum\limits_{k}}
b_{k}^{\ast}b_{k}(\frac{{\small N}}{{\small 2}}(\frac{{\small N}}{{\small 2}%
}+1)-k^{2})\nonumber\\
& -\frac{1}{4}%
{\displaystyle\sum\limits_{k}}
b_{k-2}^{\ast}b_{k}\sqrt{\frac{{\small N}}{{\small 2}}(\frac{{\small N}%
}{{\small 2}}+1)-k(k-1)}\sqrt{\frac{{\small N}}{{\small 2}}(\frac{{\small N}%
}{{\small 2}}+1)-(k-1)(k-2)}\nonumber\\
& +\frac{1}{4}\left(
\begin{array}
[c]{c}%
{\displaystyle\sum\limits_{k}}
b_{k+1}^{\ast}b_{k}\sqrt{\frac{{\small N}}{{\small 2}}(\frac{{\small N}%
}{{\small 2}}+1)-k(k+1)}\\
-%
{\displaystyle\sum\limits_{k}}
b_{k-1}^{\ast}b_{k}\sqrt{\frac{{\small N}}{{\small 2}}(\frac{{\small N}%
}{{\small 2}}+1)-k(k-1)}%
\end{array}
\right)  ^{2}\label{Eq.CovYY}%
\end{align}
and%
\begin{align}
& C_{yz}\nonumber\\
& =C(\widehat{S}_{y},\widehat{S}_{z})=C_{zy}\nonumber\\
& =\frac{1}{4i}%
{\displaystyle\sum\limits_{k}}
b_{k+1}^{\ast}b_{k}(2k+1)\sqrt{\frac{{\small N}}{{\small 2}}(\frac{{\small N}%
}{{\small 2}}+1)-k(k+1)}\nonumber\\
& -\frac{1}{4i}%
{\displaystyle\sum\limits_{k}}
b_{k-1}^{\ast}b_{k}(2k-1)\sqrt{\frac{{\small N}}{{\small 2}}(\frac{{\small N}%
}{{\small 2}}+1)-k(k-1)}\nonumber\\
& -\frac{1}{2i}%
{\displaystyle\sum\limits_{k}}
b_{k+1}^{\ast}b_{k}\sqrt{\frac{{\small N}}{{\small 2}}(\frac{{\small N}%
}{{\small 2}}+1)-k(k+1)}%
{\displaystyle\sum\limits_{k}}
b_{k}^{\ast}b_{k}\,k\label{Eq.CovYZ}\\
& +\frac{1}{2i}%
{\displaystyle\sum\limits_{k}}
b_{k-1}^{\ast}b_{k}\sqrt{\frac{{\small N}}{{\small 2}}(\frac{{\small N}%
}{{\small 2}}+1)-k(k-1)}%
{\displaystyle\sum\limits_{k}}
b_{k}^{\ast}b_{k}\,k\nonumber
\end{align}
and finally%
\begin{align}
C_{zz}  & =C(\widehat{S}_{z},\widehat{S}_{z})\nonumber\\
& =%
{\displaystyle\sum\limits_{k}}
b_{k}^{\ast}b_{k}\,k^{2}-\left(
{\displaystyle\sum\limits_{k}}
b_{k}^{\ast}b_{k}\,k\right)  ^{2}\label{Eq.CovZZ}%
\end{align}
\medskip

In terms of the new spin operators defined via the orthogonal transformation
in (\ref{Eq.NewSpinOprs}), the \textit{new covariance matrix} is given by%

\begin{equation}
C(\widehat{J}_{a},\widehat{J}_{b})=%
{\displaystyle\sum\limits_{c.d}}
M_{ac}(\theta_{p})C(\widehat{S}_{c},\widehat{S}_{d})M_{bd}(\theta_{p}%
)=\delta_{ab}\,\left\langle (\Delta\widehat{J}_{a})^{2}\right\rangle
\label{Eq.DiagonalNewCovMatrix}%
\end{equation}
where
\begin{equation}
\left[  M(\theta_{p})\right]  =\left[
\begin{tabular}
[c]{lll}%
$0$ & $0$ & $1$\\
$\sin\theta_{p}$ & $\cos\theta_{p}$ & $0$\\
$\cos\theta_{p}$ & $-\sin\theta_{p}$ & $0$%
\end{tabular}
\right]
\end{equation}
relates the new and original spin operators via
\begin{equation}
\widehat{J}_{a}=%
{\displaystyle\sum\limits_{c}}
M_{ac}(\theta_{p})\,\widehat{S}_{c}\label{Eq.NewSpinOprs0}%
\end{equation}
It turns out that the new covariance matrix is diagonal. The evaluation of the
original covariance matrix involves the following sums for $N$ large:

(a) $\sum_{k}\sqrt{(\frac{{\small N}}{{\small 2}}(\frac{{\small N}}%
{{\small 2}}+{\small 1})-k(k\pm{\small 1})}/(N+1)\doteqdot\frac{{\LARGE \pi
N}}{{\LARGE 8}}$

(b) $\sum_{k}\sqrt{(\frac{{\small N}}{{\small 2}}(\frac{{\small N}}%
{{\small 2}}+{\small 1})-k(k\pm{\small 1})}\sqrt{(\frac{{\small N}}%
{{\small 2}}(\frac{{\small N}}{{\small 2}}+{\small 1})-(k\pm1)(k\pm
{\small 2})}/(N+1)\doteqdot$ $\frac{{\LARGE N}^{2}}{{\LARGE 6}}$

(c) $\sum_{k}(\frac{{\small N}}{{\small 2}}(\frac{{\small N}}{{\small 2}%
}+{\small 1})-k^{2})/(N+1)\doteqdot$ $\frac{{\LARGE N}^{2}}{{\LARGE 6}}$

(d) $\sum_{k}(2k\pm1)\sqrt{(\frac{{\small N}}{{\small 2}}(\frac{{\small N}%
}{{\small 2}}+{\small 1})-k(k\pm{\small 1})}/(N+1)\doteqdot$ $0$. These sums
are correct to $O(N)$. This gives the new covariance matrix to $O(N)$ as
\begin{equation}
\left[  C(\widehat{J}_{a},\widehat{J}_{b})\right]  =\left[
\begin{tabular}
[c]{lll}%
$\left(  \frac{{\LARGE 1}}{{\LARGE 12}}\right)  N^{2}$ & $0$ & $0$\\
$0$ & $0$ & $0$\\
$0$ & $0$ & $\left(  \frac{{\LARGE 1}}{{\LARGE 6}}-\frac{{\LARGE \pi}^{2}%
}{{\LARGE 64}}\right)  N^{2}$%
\end{tabular}
\right] \label{Eq.CovarianceMatrix1}%
\end{equation}
showing that correct to $O(N)$ the variances for $\widehat{J}_{x},\widehat
{J}_{z}$ are large but that for $\widehat{J}_{y}$ is zero.

To determine the new covariance matrix more accurately we begin again with the
new spin operators still given by (\ref{Eq.NewSpinOprs0}) and work out the
expressions for $C(\widehat{J}_{a},\widehat{J}_{b})$ to $O(N^{0})$ when the
term is zero correct to $O(N)$. As before we find the new covariance matrix is
diagonal but now given by
\begin{equation}
\left[  C(\widehat{J}_{a},\widehat{J}_{b})\right]  =\left[
\begin{tabular}
[c]{lll}%
$\left(  \frac{{\LARGE 1}}{{\LARGE 12}}\right)  N^{2}$ & $0$ & $0$\\
$0$ & $\frac{{\LARGE 1}}{{\LARGE 8}}+\frac{1}{{\LARGE 4}}\ln N$ & $0$\\
$0$ & $0$ & $\left(  \frac{{\LARGE 1}}{{\LARGE 6}}-\frac{{\LARGE \pi}^{2}%
}{{\LARGE 64}}\right)  N^{2}$%
\end{tabular}
\right] \label{Eq.NewCovarianceMatrixFinal}%
\end{equation}
The new variance for $\widehat{J}_{y}$ is now found to be non-zero and given
by $\frac{{\LARGE 1}}{{\LARGE 8}}+\frac{1}{{\LARGE 4}}\ln N$. This requires
the following sum:

(e) $(-\sum_{k}\sqrt{(\frac{{\small N}}{{\small 2}}(\frac{{\small N}%
}{{\small 2}}+{\small 1})-k(k\pm{\small 1})}\sqrt{(\frac{{\small N}%
}{{\small 2}}(\frac{{\small N}}{{\small 2}}+{\small 1})-(k\pm1)(k\pm
{\small 2})}/2(N+1)+\sum_{k}(\frac{{\small N}}{{\small 2}}(\frac{{\small N}%
}{{\small 2}}+{\small 1})-k(k\pm1))/2(N+1))\doteqdot\frac{{\LARGE 1}%
}{{\LARGE 8}}+\frac{1}{{\LARGE 4}}\ln N$. \medskip

\subsection{Energy fluctuations}

The Josephson Hamiltonian is given by
\begin{equation}
\widehat{H}=U\widehat{S}_{z}^{2}-J\widehat{S}_{x}+\delta\widehat{S}_{z}%
\end{equation}
and it is straightforward to show that then \textit{variance} in the energy is
given by
\begin{equation}
\delta\widehat{H}^{2}=\left[
\begin{tabular}
[c]{lll}%
$U$ & $-J$ & $\delta$%
\end{tabular}
\right]  \times\left[
\begin{tabular}
[c]{lll}%
$\{\widehat{S}_{z}^{2},\widehat{S}_{z}^{2}\}$ & $\{\widehat{S}_{z}%
^{2},\widehat{S}_{x}\}$ & $\{\widehat{S}_{z}^{2},\widehat{S}_{z}\}$\\
$\{\widehat{S}_{x},\widehat{S}_{z}^{2}\}$ & $\{\widehat{S}_{x},\widehat{S}%
_{x}\}$ & $\{\widehat{S}_{x},\widehat{S}_{z}\}$\\
$\{\widehat{S}_{z},\widehat{S}_{z}^{2}\}$ & $\{\widehat{S}_{z},\widehat{S}%
_{x}\}$ & $\{\widehat{S}_{z},\widehat{S}_{z}\}$%
\end{tabular}
\right]  \times\left[
\begin{tabular}
[c]{l}%
$U$\\
$-J$\\
$\delta$%
\end{tabular}
\right] \label{Eq.EnergyVariance0}%
\end{equation}
which involves the covariances of $\widehat{S}_{z}^{2}$, $\widehat{S}_{x}$ and
$\widehat{S}_{z}$. Evaluating the covariances using certain sums given in the
previous section plus

(f) $\sum_{k}k^{2}/(N+1)\doteqdot\frac{{\LARGE N}^{2}}{{\LARGE 12}}$ (g)
$\sum_{k}k^{4}/(N+1)\doteqdot\frac{{\LARGE N}^{3}}{{\LARGE 80}}$

(h) $\sum_{k}k(k+1)\sqrt{(\frac{{\small N}}{{\small 2}}(\frac{{\small N}%
}{{\small 2}}+{\small 1})-k(k+{\small 1})}/(N+1)\doteqdot$ $\frac{{\LARGE \pi
N}^{3}}{{\LARGE 128}}$ we have correct to $O(N^{2})$%
\begin{align}
& \delta\widehat{H}^{2}\nonumber\\
& =\left[
\begin{tabular}
[c]{lll}%
$U$ & $-J$ & $\delta$%
\end{tabular}
\right]  \times\left[
\begin{tabular}
[c]{lll}%
$\frac{{\Large 1}}{{\Large 180}}N^{4}$ & $-\frac{{\Large \pi}}{{\Large 384}%
}N^{3}\cos\theta_{p}$ & $0$\\
$-\frac{{\Large \pi}}{{\Large 384}}N^{3}\cos\theta_{p}$ & $(\frac{{\Large 1}%
}{{\Large 6}}-\frac{{\Large \pi}^{2}}{{\Large 64}})N^{2}\cos^{2}\theta_{p}$ &
$0$\\
$0$ & $0$ & $\frac{{\Large 1}}{{\Large 12}}N^{2}$%
\end{tabular}
\right]  \times\left[
\begin{tabular}
[c]{l}%
$U$\\
$-J$\\
$\delta$%
\end{tabular}
\right] \nonumber\\
& =N^{2}\times\left[
\begin{tabular}
[c]{lll}%
$UN$ & $J$ & $\delta$%
\end{tabular}
\right]  \times\left[
\begin{tabular}
[c]{lll}%
$\frac{{\Large 1}}{{\Large 180}}$ & $\frac{{\Large \pi}}{{\Large 384}}%
\cos\theta_{p}$ & $0$\\
$\frac{{\Large \pi}}{{\Large 384}}\cos\theta_{p}$ & $(\frac{{\Large 1}%
}{{\Large 6}}-\frac{{\Large \pi}^{2}}{{\Large 64}})\cos^{2}\theta_{p}$ & $0$\\
$0$ & $0$ & $\frac{{\Large 1}}{{\Large 12}}$%
\end{tabular}
\right]  \times\left[
\begin{tabular}
[c]{l}%
$UN$\\
$J$\\
$\delta$%
\end{tabular}
\right] \label{Eq.EnergyVariance1}%
\end{align}
which is a \textit{quadratic form} in the quantities $UN$, $J$ and $\delta$.
That this form is positive definite can be shown by determining the eigen
values $\lambda_{1}(\theta_{p})$, $\lambda_{2}(\theta_{p})$ and $\lambda
_{3}(\theta_{p})$ of the $3x3$ matrix.

The corresponding eigenvalue equations are
\begin{equation}
\left[
\begin{tabular}
[c]{lll}%
$\frac{{\Large 1}}{{\Large 180}}$ & $\frac{{\Large \pi}}{{\Large 384}}%
\cos\theta_{p}$ & $0$\\
$\frac{{\Large \pi}}{{\Large 384}}\cos\theta_{p}$ & $(\frac{{\Large 1}%
}{{\Large 6}}-\frac{{\Large \pi}^{2}}{{\Large 64}})\cos^{2}\theta_{p}$ & $0$\\
$0$ & $0$ & $\frac{{\Large 1}}{{\Large 12}}$%
\end{tabular}
\right]  \times\left[
\begin{tabular}
[c]{l}%
$X_{1\alpha}$\\
$X_{2\alpha}$\\
$X_{3\alpha}$%
\end{tabular}
\right]  =\lambda_{\alpha}(\theta_{p})\left[
\begin{tabular}
[c]{l}%
$X_{1\alpha}$\\
$X_{2\alpha}$\\
$X_{3\alpha}$%
\end{tabular}
\right] \label{Eq.EigenMatrix0}%
\end{equation}
where the eigenvectors are orthogonal and normalised to unity $%
{\displaystyle\sum\limits_{i}}
$ $X_{i\alpha}X_{i\beta}=\delta_{\alpha\beta}$. The eigenvalues are easily
obtained as
\begin{align}
\lambda_{1}(\theta_{p})  & =\frac{1}{2}((a+b)+\sqrt{(a-b)^{2}+4c^{2}%
})\nonumber\\
\lambda_{2}(\theta_{p})  & =\frac{1}{2}((a+b)-\sqrt{(a-b)^{2}+4c^{2}%
})\nonumber\\
\lambda_{3}(\theta_{p})  & =d\label{Eq.Eigenvalues0}%
\end{align}
with $a=\frac{{\Large 1}}{{\Large 180}}$, $b=(\frac{{\Large 1}}{{\Large 6}%
}-\frac{{\Large \pi}^{2}}{{\Large 64}})\cos^{2}\theta_{p}$, $c=\frac
{{\Large \pi}}{{\Large 384}}\cos\theta_{p}$ and $d=\frac{{\Large 1}%
}{{\Large 12}}$. The eigenvalues are all real and positive, as may be seen in
Figure 10 for the non-trivial $\lambda_{1}(\theta_{p})$, $\lambda_{2}%
(\theta_{p})$.

\bigskip%
\begin{figure}
[ptb]
\begin{center}
\includegraphics[
height=3.0917in,
width=5.0548in
]%
{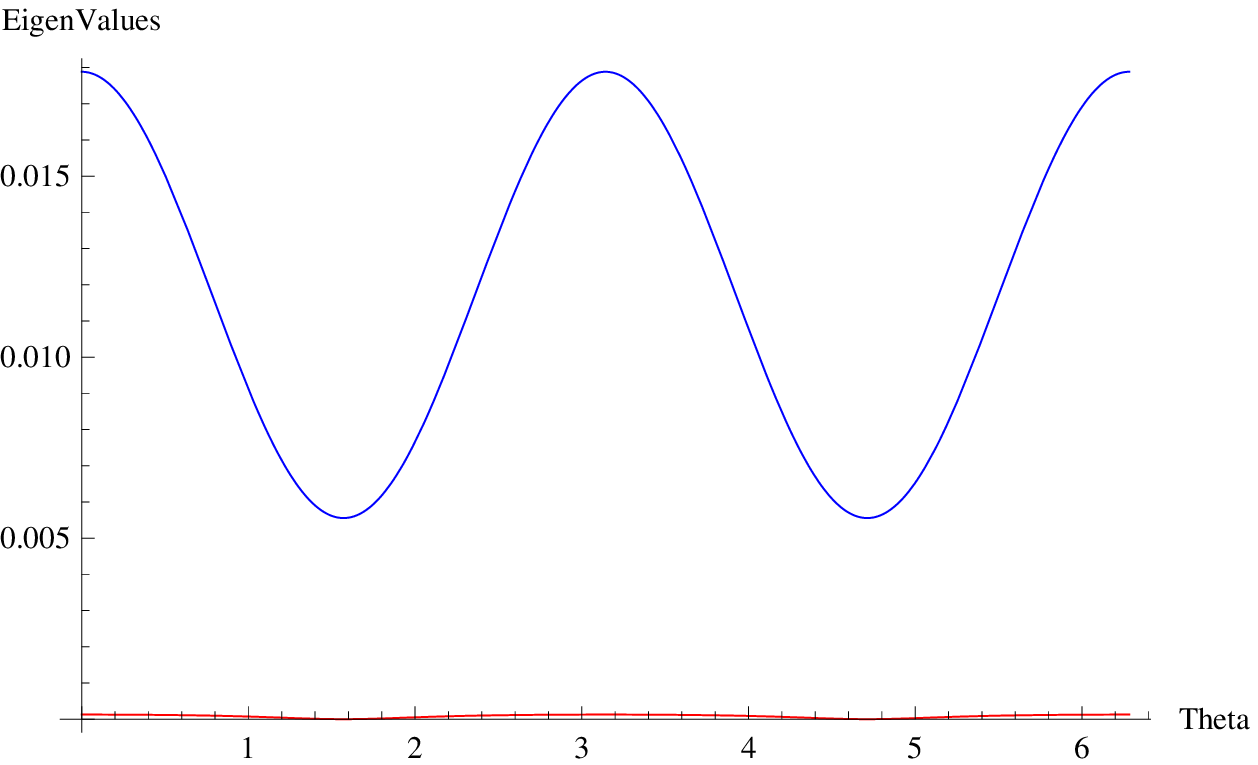}%
\end{center}
\end{figure}

\bigskip

\begin{center}
Figure 10. Eigenvalues $\lambda_{1}(\theta_{p})$ (blue curve) and $\lambda
_{2}(\theta_{p})$ (red curve) \bigskip
\end{center}

The corresponding eigenvectors are:%
\begin{align}
\left[
\begin{tabular}
[c]{l}%
$X_{11}$\\
$X_{21}$\\
$X_{31}$%
\end{tabular}
\right]   & =\frac{1}{\sqrt{(a-\lambda_{1})^{2}+c^{2}}}\left[
\begin{tabular}
[c]{l}%
$-c$\\
$(a-\lambda_{1})$\\
$0$%
\end{tabular}
\right] \nonumber\\
\left[
\begin{tabular}
[c]{l}%
$X_{12}$\\
$X_{22}$\\
$X_{32}$%
\end{tabular}
\right]   & =\frac{1}{\sqrt{(a-\lambda_{2})^{2}+c^{2}}}\left[
\begin{tabular}
[c]{l}%
$-c$\\
$(a-\lambda_{2})$\\
$0$%
\end{tabular}
\right] \nonumber\\
\left[
\begin{tabular}
[c]{l}%
$X_{13}$\\
$X_{23}$\\
$X_{33}$%
\end{tabular}
\right]   & =\left[
\begin{tabular}
[c]{l}%
$0$\\
$0$\\
$1$%
\end{tabular}
\right] \label{Eq.EigenVectors0}%
\end{align}
The eigenvector for $\lambda_{1}(\theta_{p})$ are shown in Figure 11.

\bigskip%
\begin{figure}
[ptb]
\begin{center}
\includegraphics[
height=2.9231in,
width=5.0548in
]%
{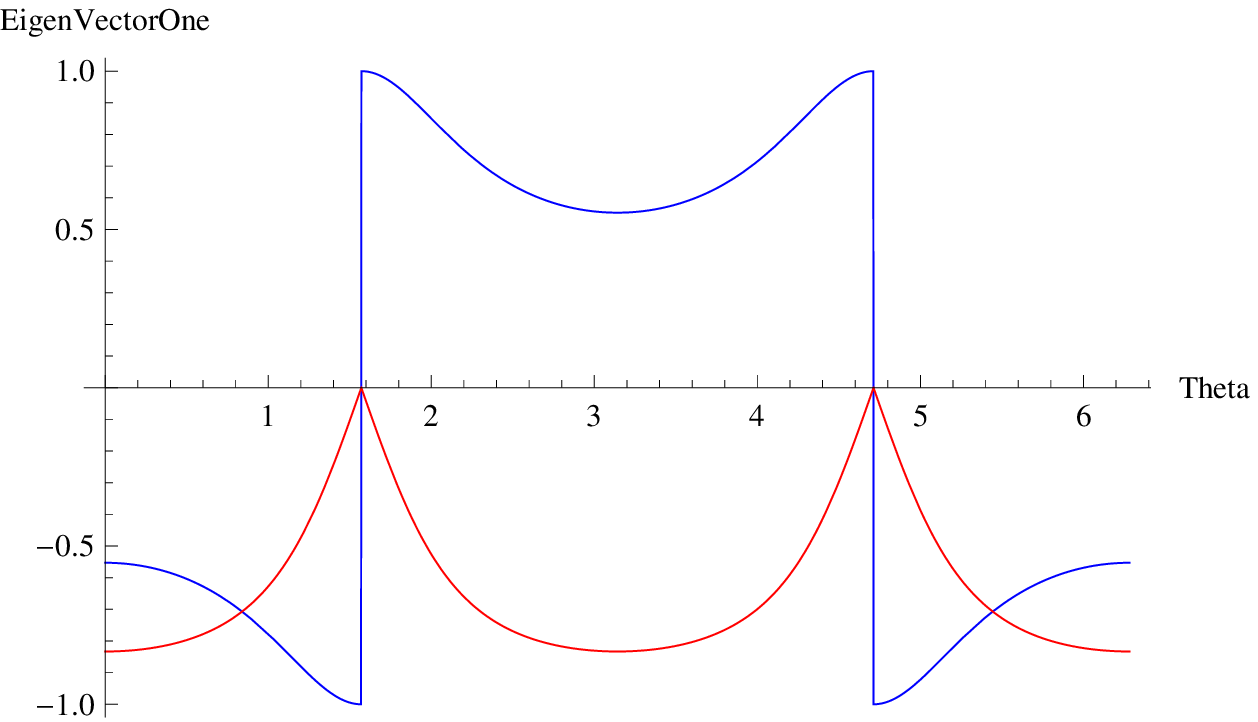}%
\end{center}
\end{figure}

\bigskip

\begin{center}
Figure 11. Eigenvector for $\lambda_{1}(\theta_{p})$. $X_{11}$ (blue curve)
and $X_{21}$ (red curve). $X_{31}=0$.\bigskip
\end{center}

For the case where
\begin{equation}
\left[
\begin{tabular}
[c]{l}%
$UN$\\
$J$\\
$\delta$%
\end{tabular}
\right]  =K\left[
\begin{tabular}
[c]{l}%
$X_{1\alpha}$\\
$X_{2\alpha}$\\
$X_{3\alpha}$%
\end{tabular}
\right] \label{Eq.SpecialJosephParam0}%
\end{equation}
where $K$ is arbitrary, it is easy to see that in this case the \textit{energy
variance} and \textit{standard deviation} are given by%
\begin{align}
\left(  \delta\widehat{H}^{2}\right)  _{\alpha}  & =N^{2}K^{2}\lambda_{\alpha
}(\theta_{p})\nonumber\\
\sqrt{\left(  \delta\widehat{H}^{2}\right)  _{\alpha}}  & =NK\sqrt
{\lambda_{\alpha}(\theta_{p})}\label{Eq.EnergyFluctnSpecialParam0}%
\end{align}
The \textit{average energy} in this case is
\begin{align}
\left\langle \widehat{H}\right\rangle  & \doteqdot N\left(  UN\frac{1}%
{12}-J\cos\theta_{p}\frac{\pi}{8}\right) \nonumber\\
\left\langle \widehat{H}\right\rangle _{\alpha}  & \doteqdot NK\left(
X_{1\alpha}\frac{1}{12}-X_{2\alpha}\cos\theta_{p}\frac{\pi}{8}\right)
\label{Eq.AverEnergySpecialParam0}%
\end{align}
so the \textit{relative energy fluctuation} is
\begin{equation}
\frac{\sqrt{\left(  \delta\widehat{H}^{2}\right)  _{\alpha}}}{\left\vert
\left\langle \widehat{H}\right\rangle _{\alpha}\right\vert }=\frac
{\sqrt{\lambda_{\alpha}(\theta_{p})}}{\left\vert \left(  X_{1\alpha}\frac
{1}{12}-X_{2\alpha}\cos\theta_{p}\frac{\pi}{8}\right)  \right\vert
}\label{Eq.RelativeEnergyFluctnSpecialParam0}%
\end{equation}
This expression only applies to the $\lambda_{1}(\theta_{p})$, $\lambda
_{2}(\theta_{p})$ eigenvalues, since for $\lambda_{3}(\theta_{p})$ we have
$\left\langle \widehat{H}\right\rangle _{3}=0$. \pagebreak


\bigskip

\begin{thebibliography}{99}                                                                                               %
\bibitem {Barnett93a}S. M. Barnett and B. J. Dalton, Phys. Scripta
\textbf{T48}, 13 (1993).

\bibitem {Barnett89a}S. M. Barnett and D. T. Pegg, Phys. Rev. A \textbf{39},
1665 (1989).

\bibitem {Schleich92a}W. Schleich, A. Bandilla and H. Paul, Phys. Rev. A
\textbf{45}, 6652 (1992).

\bibitem {Smithey93a}D. T. Smithey, M. Beck, J. Cooper and M. G. Raymer, Phys.
Scripta \textbf{T48}, 35 (1993).

\bibitem {Shapiro89a}J. H. Shapiro, S. R. Shepard and N. C. Wong, Phys. Rev.
Letts. \textbf{62}, 2377 (1989).

\bibitem {Noh91a}J. W. Noh, A. Fougeres and L. Mandel, Phys. Rev. Letts.
\textbf{67}, 1426 (1991).

\bibitem {Pegg97a}D. T. Pegg and S. M. Barnett, J. Mod. Opt., \textbf{44}, 225 (1997).

\bibitem {Dirac27a}P. A. M. Dirac, Proc. Roy. Soc. A, \textbf{114}, 243 (1927).

\bibitem {Susskind64a}L. Susskind and J. Glogower, Physics, \textbf{1}, 49 (1964).

\bibitem {Nieto93a}M. M. Nieto, Phys. Scripta \textbf{T48}, 5 (1993).

\bibitem {Vaccaro95a}J. Vaccaro, Phys. Rev. A \textbf{51}, 3309 (1995).

\bibitem {Luis93a}A. Luis and L. L. Sanchez-Soto, Phys. Rev. A \textbf{48},
4702 (1993).

\bibitem {Luis96a}A. Luis and L. L. Sanchez-Soto, Phys. Rev. A \textbf{53},
495 (1996).

\bibitem {Menotti01a}C. Menotti, J. R. Anglin, J. I. Cirac and P. Zoller,
Phys. Rev. A \textbf{63}, 023601 (2001).

\bibitem {Barnett90a}S. M. Barnett and D. T. Pegg, Phys. Rev. A \textbf{42},
6713 (1990).

\bibitem {Pegg95a}D. T. Pegg and J. A. Vaccaro, Phys. Rev. A \textbf{51}, 859 (1995).

\bibitem {Luis95a}A. Luis and L. L. Sanchez-Soto, Phys. Rev. A \textbf{51},
861 (1995).

\bibitem {Li09a}Y. Li, P. Treutlein, J. Reichel and A. Sinatra, Eur. Phys. J.
B, \textbf{68}, 365, (2009).

\bibitem {Grond10a}J. Grond, U. Hohenester, I. Mazets and J. Schmiedmayer, New
J. Phys. \textbf{12}, 065036 (2010).

\bibitem {Leggett01a}A. J. Leggett, Rev. Mod. Phys. \textbf{73}, 307 (2001).

\bibitem {Vogel91a}W. Vogel and W. Schleich, Phys. Rev. A \textbf{44}, 7642 (1991).

\bibitem {Amico08a}L. Amico, R. Fazio, A. Osterloh and V. Vedral, Rev. Mod.
Phys. \textbf{80}, 517 (2008).

\bibitem {Hines03a}A. P. Hines, R. H. McKenzie and G. J. Milburn, Phys. Rev. A
\textbf{67}, 013609 (2003).

\bibitem {Dalton07a}B. J. Dalton. J. Mod. Opt. \textbf{54}, 615 (2007).

\bibitem {Dalton11a}B. J. Dalton, Ann. Phys. \textbf{326}, 668 (2011).

\bibitem {Kitagawa93a}M. Kitagawa and M. Ueda, Phys. Rev. A \textbf{47}, 5138 (1993).

\bibitem {Jaaskelainen06a}M. Jaaskelainen and P. Meystre, Phys. Rev. A
\textbf{73}, 013602 (2006).

\bibitem {Bouyer97a}P. Bouyer and M. A. Kasevich, Phys. Rev. A \textbf{56},
R1083 (1997).

\bibitem {Bach04a}R. Bach and K. Rzazewski, Phys. Rev. Letts. \textbf{92},
200401 (2004).

\bibitem {Bach04b}R. Bach and K. Rzazewski, Phys. Rev. A \textbf{70}, 063622 (2004).

\bibitem {Dunningham04a}J. A. Dunningham and K. Burnett, Phys. Rev. A
\textbf{70}, 033601 (2004).

\bibitem {Wright96a}E. M. Wright, D. F. Walls and J. C. Garrison, Phys. Rev.
Letts. \textbf{77}, 2158 (1996).
\end{thebibliography}
\end{document}